\documentclass[journal]{IEEEtran}

\usepackage{bbm}
\usepackage{cite}
\usepackage{amsmath}
\usepackage{amssymb}
\usepackage{amsfonts}
\usepackage{graphicx}
\usepackage{epstopdf}
\usepackage{algorithm}
\usepackage{algorithmic}
\usepackage{epsfig}
\usepackage{epstopdf}
\usepackage{subfigure}
\usepackage{url}

\usepackage{pifont}
\usepackage{psfrag}
\usepackage{times}
\usepackage{booktabs}
\usepackage[T1]{fontenc}
\usepackage{moreverb}
\usepackage{color}
\usepackage{setspace}
\usepackage{mathrsfs}
\usepackage{array}
\usepackage{txfonts}
\usepackage{makecell}
\usepackage{bm}
\newtheorem{lemma}{Lemma}

\newtheorem{remark}{Remark}
\newtheorem{proposition}{Proposition}

\usepackage{fancyhdr}

\begin{document}
\makeatletter
\def\ps@IEEEtitlepagestyle{%
	\def\@oddhead{%
		\scriptsize
		\parbox{\textwidth}{%
			\centering
This article has been published in IEEE Transactions on Wireless Communications. This is the author's version which has not been fully edited and content may change prior to final publication. Citation information: DOI 10.1109/TWC.2024.3477750.
		}%
	}%
	\def\@evenhead{\@oddhead}
	\def\@oddfoot{}%
	\def\@evenfoot{\@oddfoot}%
}
\makeatother
	\title{Holographic Integrated Data and Energy Transfer}
	\author{ Qingxiao Huang,~\IEEEmembership{Graduate Student Member,~IEEE}, Jie Hu,~\IEEEmembership{Senior Member,~IEEE}, Yizhe Zhao,~\IEEEmembership{Member,~IEEE} and Kun~Yang,~\IEEEmembership{Fellow,~IEEE}
		\thanks{This work was supported in part by the Natural Science Foundation of China (NSFC) under Grant 62201123; in part by the MOST Major Research and Development Project under Grant 2021YFB2900204; in part by the Natural Science Foundation of China under Grant 62431002; in part by the Young Elite Scientists Sponsorship Program by CAST under Grant 2023QNRC001; in part by the Natural Science Foundation of Sichuan (NSFSC) under Grant 2024NSFSC1417; in part by the Stable Supporting Fund of National Key Laboratory of Underwater Acoustic Technology under Grant JCKYS2023604SSJS005; in part by the Natural Science Foundation of China under Grant 62132004; in part by the Municipal Government of Quzhou under Grant 2023D005; and in part by the China Postdoctoral Science Foundation under Grant 2022TQ0056. {\it (Corresponding author:	Jie Hu.)}}
			\thanks{Qingxiao Huang, Jie Hu and Yizhe Zhao are with the School of Information and Communication Engineering, University of Electronic Science and Technology of China, Chengdu, 611731, China, email: qxhuang@std.uestc.edu.cn; hujie@uestc.edu.cn and yzzhao@uestc.edu.cn. }
	\thanks{Kun Yang is with State Key Laboratory of Novel Software Technology, Nanjing University, Nanjing 210008, China, and School of Intelligent Software and Engineering, Nanjing University (Suzhou Campus), Suzhou, 215163, China, and School of Computer Science and Electronic Engineering, University of Essex, U.K., email: kyang@ieee.org.}	
}
	\maketitle
	\begin{abstract}
Thanks to the application of metamaterials, holographic multiple-input multiple-output (H-MIMO) is expected to achieve a higher spatial diversity gain by enabling the ability to generate any current distribution on the surface. With the aid of electromagnetic (EM) manipulation capability of H-MIMO, integrated data and energy transfer (IDET) system can fully exploits the EM channel to realize energy focusing and eliminate inter-user interference, which yields the concept of holographic IDET (H-IDET). In this paper, we invetigate the beamforming designs for H-IDET systems, where the sum-rate of data users (DUs) are maximized by guaranteeing the energy harvesting requirements of energy users (EUs). In order to solve the non-convex functional programming, a block coordinate descent (BCD) based scheme is proposed, wherein the Fourier transform and the equivalence between the signal-to-interference-plus-noise ratio (SINR) and the mean-square error (MSE) are also conceived, followed by the successive convex approximation (SCA) and an initialization scheme to enhance robustness. Numerical results illustrate that our proposed H-IDET scheme outperforms benchmark schemes, especially the one adopting traditional discrete antennas. Besides, the near-field focusing using EM channel model achieves better performance compared to that using the traditional channel model, especially for WET where the EUs are usually close to the transmitter. 
	\end{abstract}


\section{Introduction}
\subsection{Background}
In the era of next Internet of Things (IoT), ubiquitous intelligent connections are establishing between people and things. Deploying millions or even billions of IoT devices having low hardware size, complexity and power consumption into smart cities will readily breeds some emerging applications such as intelligent transportation, virtual reality and cloud computing \cite{10122600}. However, it is posing a challenge for the energy requirement of all the IoT devices, where manually replacing or charging batteries requires much high maintenance costs and even brings about some safety hazards. In the Ambient IoT, the low-power devices are expected to harvest energy by light, motion, heat and especially radio frequency (RF) signals \cite{RP-234058}. By utilizing existing transmitters such as cellular base stations, TV towers, and Wi-Fi access points, RF based wireless energy transfer (WET) has the advantages of better controllability and stability \cite{8421584}. By coordinating with wireless data transfer (WDT), integrated data and energy transfer (IDET) is a promising technology in the future to provide both energy self-sustaining and data communication services towards low-power devices in IoT \cite{9261955}. 

Recently, massive multiple-input-multiple-output (MIMO) technology has been widely studied to achieve a high spatial gain and reduce interference among users. Due to the implementation difficulties of fabricating subwavelength sized antennas and the mutual coupling that occurs between closely spaced antennas, the antennas in classical MIMO systems are usually half-wavelength spacing \cite{9724245}. Thanks to the advances of metasurface, each metamaterial element of the array is subwavelength sized and separated. With the aid of the densely spaced metamaterial elements, holographic MIMO (H-MIMO) is able to form a quasi-continuous electromagnetic (EM) surface and is expected to have the strong ability to arbitrarily manipulate the current distribution on the surface, which can fully exploit the propagation characteristics arised by an EM channel \cite{9136592}. 

Moreover, thanks to the ability to manipulate the EM field, H-MIMO can not only ultimately approaching the communication limit but also achieve unprecedented concentration of EM energy, which is beneficial to both the WDT and WET \cite{9627147}. With the aid of the propagation characteristics of the EM channel, the unprecedented energy concentration and multiplexing are realized for EUs, while the inter-user interference is well eliminated for DUs, which yields the concept of Holographic IDET (H-IDET).

\subsection{Related works}
In the multi-user IDET systems, the interference signals are usually harmful for WDT but can improve the WET performance. Therefore, the transceivers should be carefully designed to facilitate efficient IDET. Specifically, Xiong \emph{et al.} \cite{7934322} investigated a broadcasting IDET system and analyzed the tradeoff between energy transfer versus information rate under nonlinear energy harvesting model. Yue \emph{et al.} \cite{9145622} studied a millimeter wave IDET multicast system, where the data rate was maximized by satisfying wireless charging requirement with the aid of hybrid transceiver design. Large-scale antenna array are used in IDET to alleviate the inherent power attenuation caused by path loss in the channel. Gao \emph{et al.} \cite{9849458} investigated an active intelligent reflecting surface (IRS) aided IDET system, where the beamforming designs were obtained by iterative algorithms, in order to maximize the sum-rate of DUs and sum-power harvested by EUs. Zhang \emph{et al.} \cite{10032267} considered an IRS assisted IDET system, where a deep reinforcement learning-based approach is proposed by utilizing proximal policy optimization framework, in order to maximize the energy efficiency.

Moreover, there are lots of works studying H-MIMO. The metamaterial element of H-MIMO should satisfy the frequency response of Lorentzian form, which leads to different hardware structures \cite{PhysRevApplied.8.054048}. Precisely, Deng \emph{et al.} \cite{10163760} proposed a hardware design for H-MIMO to achieve amplitude-controlled beamforming, where a prototype was implemented to support real-time video transmission. Based on the amplitude-only-controlled hardware structure, they investigated the H-MIMO for cell communication systems \cite{9696209}, low-earth-orbit satellite systems \cite{9848831} and integrate sensing and communication (ISAC) systems \cite{9724245}. Besides, different from amplitude-control, the elements of the phase-only-controlled H-MIMO has Lorentzian-constrained phase model while keeping the amplitude constant, which can achieve considerable performance \cite{8756024,9738442}. In addition, J. An \emph{et al.} \cite{10158690} investigated a stacked intelligent metasurfaces (SIM) aided H-MIMO communication system, where the novel SIM can accomplish transmit precoding and receiver combining in the native EM wave regime without excessive RF chains. 

Furthermore, in order to explore the performance limits enabled by H-MIMO, EM theories should be relied upon for the physical antenna design, communication capacity analysis and channel modelling \cite{10130641,10130647,10130638}. Specifically, Badawe \emph{et al.} \cite{nature_Metasurface} proposed a practical physical architecture for the H-MIMO, where experimental results of the fabricated prototype matche well with numerical simulations. Besides, Hu \emph{et al.} \cite{8319526} considered a communication system with a H-MIMO transmitter and several single-antenna receivers, where the normalized capacity and Degree of Freedom (DoF) measured per unit surface or per unit volume  was analyzed. Moreover, Dardari \emph{et al.} \cite{9139337} investigated the communication between two H-MIMO transceivers and the spatial DoF is derived in analytical expression.  Furthermore, Pizzo \emph{et al.} \cite{9724113} investigated the channel model of H-MIMO in scattering environments, where the model  is further applied to evaluate the capacity of H-MIMO communication. In recent years, the EM based H-MIMO transceiver designs for communication systems are up to date. Specifically, Sanguinetti \emph{et al.} \cite{9906802} investigated the line-of-sight communication between two H-MIMO transceivers in the near-field. By utilizing Fourier basis functions, a multiplexing scheme was designed to generate orthogonal beamformers. Besides, Zhang \emph{et al.} \cite{10158997} considered a multi-user H-MIMO system, where a pattern-division multiplexing in the wavenumber domain was proposed to support general communication scenarios.

By utilizing the EM focusing capability of H-MIMO, efficient near-field WDT and WET were achieved, respectively. Specifically, zhang \emph{et al.} \cite{9738442} investigated a near-field WDT system. The near-field focusing is achieved by fully-digital MIMO, hybrid digital-analog MIMO and H-MIMO, while the H-MIMO performs the best among three structures with a given aperture in the single-user situation. Moreover, Sha \emph{et al.} \cite{10146329} investigated a H-MIMO assisted WDT system, where a three-dimensional near-field beamforming design is proposed to  significantly reduce codebook-size and can be compatible with the fifth-generation new-radio (5G-NR) system. Besides, Wei \emph{et al.} \cite{10103817} considered a tri-polarized H-MIMO aided near-field WDT system. Based on the constructed channel model, two beamforming schemes were proposed to mitigate the interference and improve spectral efficiency. Wu \emph{et al.} \cite{10086987} investigated a H-MIMO assisted multi-user WET system, where the phase and amplitude of H-MIMO were jointly optimized to generate multifocal beams, in order to improve the transmission efficiency. Zhang \emph{et al.} \cite{8793087} proposed a design of dual-polarized reflective H-MIMO, where near-field focus was realized in a WET system. Moreover, Zhao \emph{et al.} \cite{10310454} designed a dual-frequency H-MIMO for IDET systems, where each frequency corresponds to a different topology structure in the same aperture.

\subsection{Motivation and Contributions}
However, the existing researches have the following drawbacks:
\begin{itemize}
	\item The traditional channel for MIMO systems is a discretization of the EM fields, which mismatches the continuous nature of practical three-dimensional (3D) EM fields. Besides, approximation operations, such as plane wave and Fresnel approximation, are used in traditional channel models. Therefore, there is still a gap between the traditional channel model and the practical EM channel.
	\item The Friss formula is usually used to calculate the harvested energy in WET. However, it fails in the near-field range since the formula is under the far-field approximation. In addition, the effect of the beamforming gain and antenna polarization should be further explored in the energy harvesting model.
	\item Typically, EM theory is used in the fields of channel modeling, capacity calculation, DoF analysis and communication mode design. However, in H-IDET systems, there is a paucity of treatises on the tradeoff between data and energy transfer based on EM theory. Furthermore, researches on the H-IDET in near-field scenarios remains vacant.
\end{itemize}

Against this background, our novel contributions are summarized as follows:
\begin{itemize}
	\item We innovatively investigate a EM theory based H-IDET system, where Poynting's theorem is relied upon for the WET model. A continuous aperture H-IDET transmitter with three polarizations is considered to be compatible with the EM channel for achieving a better IDET performance.
	\item In the single EU system, the optimal beamforming for WET is obtained in closed form, where the beam can be focused at the position of different distance  but having the same angle. Furthermore, by using the H-MIMO and EM channel model, we extend the lower bound of the applicable range of traditional near-field focusing.
	\item  In the multi-user system, we study the holographic beamforming design for maximizing the sum rate of all the DUs by guaranteeing the energy harvesting performance of EUs. A block coordinate descent (BCD) based algorithm is proposed for obtaining the optimal solution, wherein the Fourier transform and successive convex approximation (SCA) are also conceived, followed by initialization schemes to enhance robustness. 
	\item  Numerical results evaluates the performance of our H-IDET system, which also demonstrate that, the H-IDET scheme outperforms the benchmarks on both near-field focusing and spatial filtering. Besides, the near-field focusing using EM channel model achieves better performance compared to that using the traditional channel model, especially for WET where the EUs are usually close to the transmitter.
\end{itemize}

The rest of the paper is organised as follows: In Section II, we describe the transceiver architecture and electromagnetic channel model \uppercase\expandafter{\romannumeral2}, while the near-field focus for a single EU system and the multi-user H-IDET system design are investigated in Section IV. Then, our numerical results are presented in Section V, followed by the conclusions in Section VI.

Notation: $(\cdot)^H$ denotes conjugate transpose operations; $\mathbb{E} \{ \cdot \}$ represents the statistical expectation; $|a|$ and $||\mathbf{a}||$ are the magnitude and norm of a scalar $a$ and vector $\mathbf{a}$, respectively; $||\mathbf{A}||$ denotes the Frobenius norm of the matrix $\mathbf{A}$; $\mathbf{A}(i,j)$ represents the specific element in the $i$-th row and $j$-th column of $\mathbf{A}$; `:=' is signified as `is defined to equal'; $ \mathfrak{R}(\cdot) $ denotes the real part of its argument.

\section{Tansceiver and Channel Model}
\begin{figure*}[h]
	\centering
	\includegraphics [width=172mm]{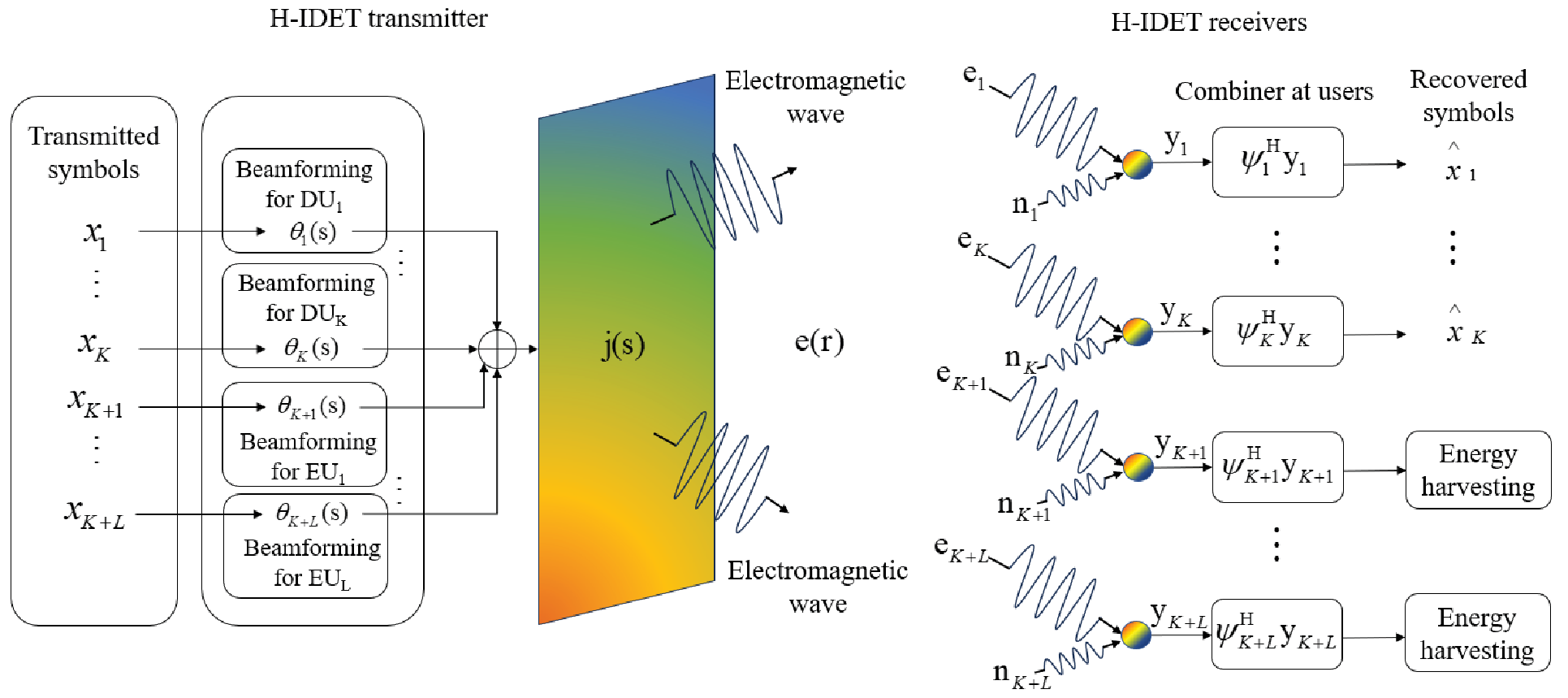}\\
	\caption{H-IDET system} 
	\label{transceiver}
\end{figure*}
\subsection{H-IDET Transmitter}
The transmitter architecture of H-IDET is portrayed in Fig. \ref{transceiver} , which works in a 3-D infinite and homogeneous medium with aperture $ {\mathcal{S}}_{\text{T}} $ of area $ A_{\text{T}}=|{\mathcal{S}}_{\text{T}}| $. It is assumed that the H-IDET transmitter has continuous antenna aperture and can synthesize any current distribution on its surface. Without loss of generality, we assume that the system works in the narrow band, while the current density distribution (amplitude and phase) at position $ \mathbf{s} :=(s_x,s_y,s_z) \in \mathbb{R}^{3\times 1}$ is denoted as $ \mathbf{j(s)} \in \mathbb{C}^{3\times 1}, \mathbf{s} \in {\mathcal{S}}_{\text{T}}$. A downlink H-IDET system is studied, where the transmitter serves $ K $ data users (DUs) and $ L $ energy users (EUs), which are all equipped with holographic receivers simultaneously. Let $ \mathcal{D} $ and $ \mathcal{E} $ represent the set of DUs and EUs, respectively. Then, the set of all users is given by $ \mathcal{U}=\mathcal{D} \cup \mathcal{E} $. The symbols transmitted to the users are denoted as $ \mathbf{x}\stackrel{\vartriangle}{=} [x_1,\dots,x_K,x_{K+1},\dots,x_{K+L}]^T \in \mathbb{C}^{K+L} $, which are assumed as Gaussian distributed\footnote{The channel capacity is able to be achieved with the Gaussian distributed transmit signals, which reveals the upper-bound performance of the practical H-IDET system. Note that the Gaussian distributed signals can be asymptotically generated by appropriately designing the modulation and coding schemes for WDT and WET. Specifically, the uncorrelatedness of different WDT streams can be assured by assigning orthogonal codes to different DUs \cite{Golomb_Gong_2005}. Moreover, by using pseudo random codes to generate the WET waveforms, they are assured to be uncorrelated with other WET waveforms or with the WDT streams \cite{9551937}.} and have normalized power, \textit{i.e.}, $ \mathbb{E}\{\mathbf{x}\mathbf{x}^H\}= \mathbf{I}_{K+L} $. Similar to the conventional muti-user beamforming, $ \bm{\theta}_k(\mathbf{s}), {\mathbf{s}} \in {{\mathcal{S}}_{\text{T}}} $ is denoted as the beamformer that H-IDET uses to transmit the symbol $ x_k $ to the user $ k \in \mathcal{U}$. For the simplicity, we assume that beamformers $ \{\bm{\theta}_k(\mathbf{s})\}^{K}_{k=1} $ are linearly superposed by the H-IDET transmitter, then the current distribution $ \mathbf{j(s)} $ can be modeled as \cite{10158997}
\begin{equation}\label{eq:1}
	\mathbf{j(s)}=\sum_{k=1}^{K+L} \bm{\theta}_k(\mathbf{s})x_k,
\end{equation}
where beamformer $ \bm{\theta}_k(\mathbf{s}) $ is the component of current density for modulating the symbol $ x_k $.
\begin{remark}
	The Beamformer for each user is essentially a different current distribution over the whole aperture, which the signal of each user is modulated on. For simplicity, it is assumed that the transmitter employs linear superposition to combine multiple information-carrying current distribution \cite{10158997}. In practice, there are practical hardware architectures that are consistent with our assumptions. For example, in \cite{9681843} the signal is processed in baseband, then linear superposed and transmitted to the whole aperture by the RF chain and feeds. Therefore, each user can utilize the whole transmitter aperture by the beamformer, and the transmitter is not pre-divided into multiple groups. However, in order to obtain better H-IDET performance, the beamformer of each DU tends to utilize the aperture at different positions and has different phases. The beamformer of each EU tends to have the same current distribution.
\end{remark}

\subsection{Electromagnetic Channel}
At the H-IDET receiver side, an information-carrying electric field $ \mathbf{e(r)} $ at position $ \mathbf{r} :=(r_x,r_y,r_z) \in \mathbb{R}^{3\times 1}$ is produced by the current density distribution $ \mathbf{j(s)} $ on the H-IDET transmit aperture, which satisfies the inhomogeneous Helmholtz wave equation \cite{9906802}
\begin{equation} \label{eq:2}
	\nabla \times \nabla \times {\mathbf{e}}({\mathbf{r}})-\kappa ^{2} {\mathbf{e}}({\mathbf{r}})= \mathsf {i}\kappa Z_{0} {\mathbf{j}}(\boldsymbol{ { {\mathbf{r}}}}),
\end{equation}
where $ \kappa=2\pi / \lambda $ is the spatial wavenumber, $ \lambda $ is the wavelength, $ \mathsf {i} $ is the imaginary unit and $ Z_0 $ is the characteristic impedance of spatial medium, which is 376.73 Ω in the free space. With the aid of the Green's method \cite{4685903} for solving (\ref{eq:2}), the electric field $ \mathbf{e(r)} $ is given by 
\begin{equation} \label{eq:3}
	\mathbf {e}(\mathbf {r})=\int _{\cal S_{\mathrm{ T}}} \mathbf {G}(\mathbf {r}, \mathbf {s}) \mathbf {j}(\mathbf {s}) \mathrm {d} \mathbf {s}, 
\end{equation}
where Green function $ \mathbf {G}(r,s) $ can be regarded as the channel matrix in holographic systems and is determined by the specific wireless environment. In an unbounded and homogeneous mediums medium, $ \mathbf {G}(r,s) $ is expressed as \cite{1386525}
\begin{align} \label{eq:4}
	{\mathbf{G}}({\mathbf{r}}, {\mathbf{s}})&= \frac {\mathsf {i}\kappa Z_0}{4 \pi }\frac {e^{ \mathsf {i}\kappa \|  {\mathbf{p}}\|}}{\| {\mathbf{p}}\|} \bigg[ \left(\mathbf {I_3} - \hat{\mathbf{p}} \hat {\mathbf{p}}^H \right) \nonumber \\&+\frac{\mathsf {i}}{\kappa\| {\mathbf{p}}\|}\left(\mathbf {I_3} - 3\hat{\mathbf{p}} \hat {\mathbf{p}}^H \right)+\frac{\mathsf {1}}{(\kappa\| {\mathbf{p}}\|)^2}\left(\mathbf {I_3} - 3\hat{\mathbf{p}} \hat {\mathbf{p}}^H \right)\bigg] .
\end{align}
where $ \mathbf{p}=\mathbf{r}-\mathbf{s} $. In EM theory, the three terms in $ \mathbf {G}(r,s) $ from left to right correspond to the far-field, the middle-field, and the near-field radiation, respectively. When $ \| \mathbf{r}-\mathbf{s} \| \gg \lambda $, the second and third terms can be neglected and an EM far-field approximation of $ \mathbf {G}(r,s) $ is then obtained. \footnote{The latter two terms correspond to the evanescent wave, an electromagnetic wave propagating on the surface of a medium, whose amplitude decays rapidly with distance. On the contrast, only the first term corresponds to the radiated field since it falls off inversely as the distance.} Traditionally, in the antenna theory, the first term can be further divided into radiation far-field and radiation near-field according to the border radius $ r=2D^2/\lambda $, wherein $ D $ is the maximum projection length of the antenna aperture in the corresponding direction.

\subsection{Holographic Receiver for DUs }
In our H-IDET system, all the users are located in the radiated field region. The holographic area at each user is far less than that of transmitter, i.e., $ A_{\text{R}}= \frac{\lambda^2}{4\pi} \ll A_{\text{T}} $. Therefore, each user can be approximated as a point in the 3-D space \cite{8319526}. 

The location of the $ k $-th DU is denoted as $ \mathbf{r}_k \in \mathbb{R}^{3\times 1} $. It is assumed that the three polarizations of EM waves can be sensed by a tri-polarization antenna. According to (\ref{eq:1}) and (\ref{eq:3}), the EM wave received by the $ k $-th DU is denoted as
\begin{align} \label{eq:5}
	\mathbf{y}_k = x_k \int_{\mathcal{S}_{\mathrm{ T}}} \mathbf{G}_k(\mathbf{s}) \bm{\theta}_k(\mathbf{s})\mathrm {d}\mathbf{s}+\sum^{K+L}_{j=1,j \neq k} x_j \int_{\mathcal{S}_{\mathrm{ T}}} \mathbf{G}_k(\mathbf{s})\bm{\theta}_j(\mathbf{s})\mathrm {d}\mathbf{s}+\mathbf{n}_k,
\end{align}
where $ \mathbf{G}_k(\mathbf{s}):= \mathbf{G}(\mathbf{r}_k,\mathbf{s})$ and $ \mathbf{n}_k $ accounts for the EM noise at the $ k $-th DU. Under the
isotropic condition, $ \mathbf{n}_k $ for all $ k \in \mathcal{D} $ can be modeled as mutually independent Gaussian distribution having the zero mean and the variance $ \sigma^2 \mathbf{I}_3 $ \cite{10158997}.

By conceiving the polarization combiner $\boldsymbol{\psi}_k\in\mathbb{C}^{3\times1}$ at the DU, the recovered signal can be expressed as $ \boldsymbol{\psi}_k^H\mathbf{y}_k $. Then, the SINR of the $ k $-th DU is given by
\begin{equation}\label{eq:sinr}
	\Gamma_k= \dfrac{\left| \bm{\psi}_k^H \int_{\mathcal{S}_{\mathrm{ T}}} \mathbf{G}_k(\mathbf{s}) \bm{\theta}_k(\mathbf{s})\mathrm {d}\mathbf{s}\right|^2 }{\sum^{K+L}_{j=1,j \neq k} \left| \bm{\psi}_k^H \int_{\mathcal{S}_{\mathrm{ T}}} \mathbf{G}_k(\mathbf{s})\bm{\theta}_j(\mathbf{s})\mathrm {d}\mathbf{s}\right|^2+\sigma^2 \bm{\psi}_k^H\bm{\psi}_k}     ,
\end{equation}
The achievable data rate of the $ k $-th DU can be further expressed as
\begin{equation}
	R_k= \log_2\left(1 + \Gamma_k  \right)  .
\end{equation}


\subsection{Holographic Receiver for EUs}	
In general, the power density at the EUs can be expressed in terms of the time-averaged magnitude of the Poynting vector, which can be caculated by:
\begin{align}
	S(\mathbf{r})=\frac{1}{2Z}\mathbf{e}^{H}_l(\mathbf{r})\mathbf{e}_l(\mathbf{r}) &=\frac{1}{2Z}\left\|  \sum^{K+L}_{j=1} x_j \int_{\mathcal{S}_{\mathrm{T}}} \mathbf{G}_l(\mathbf{s}) \bm{\theta}_j(\mathbf{s})\mathrm{d}\mathbf{s} \right\| ^2 \nonumber
\end{align}
\begin{align}
	&\overset{\text{(a)}}{=}\frac{1}{2Z}\sum^{K+L}_{j=1}\left\|  \int_{\mathcal{S}_{\mathrm{T}}} \mathbf{G}_l(\mathbf{s})\bm{\theta}_j(\mathbf{s})\mathrm{d}\mathbf{s} \right\| ^2,
\end{align}
where $ Z $ is the wave impedance of the receiving H-IDET, (a) is realized by utilizing $ \mathbb{E}\{\mathbf{x}\mathbf{x}^H\}= \mathbf{I}_{K+L} $. According to the Poynting's theorem, the total power flows into the receiving holographic area of the $ l $-th EU can be derived as 
\begin{align} \label{eq:6}
	P_l &= \int_{ \mathcal{S}_{\mathrm{R}}} \mathbf{S}(\mathbf{r})\cdot \hat{\boldsymbol{\eta}} \mathrm{d}\mathbf{s}  =\frac{A_{\text{R}}\cos(\phi) }{2Z}\sum^{K+L}_{j=1}\left\|  \int_{\mathcal{S}_{\mathrm{T}}} \mathbf{G}_l(\mathbf{s})\bm{\theta}_j(\mathbf{s})\mathrm{d}\mathbf{s} \right\| ^2, 
\end{align}
where $ \hat{\boldsymbol{\eta}} \in \mathbb{R}^{3\times 1} $ denotes the normal vector of receiving surface, $ \phi \in \left [0,90^\circ \right ] $  denotes the included angle between $ \hat{\boldsymbol{\eta}} $ and Poynting vector at location $ \mathbf{r} $.

Moreover, the non-linear energy harvesting model is conceived in our work, where the energy harvesting power at the $ l $-th EU can be modeled as \cite{7934322}
\begin{equation}\label{eq:8}
	\Xi(P_l)=\dfrac{M}{X(1+\exp(-a(P_l-b)))}-Y
\end{equation}
where $ X=\frac{\exp(ab)}{1+\exp(ab)} $ and $ Y=\frac{M}{\exp(ab)} $. Besides, $ M $ represents the maximum output directcurrent power when the circuit is saturated, while $ a $ and $ b $ are constants which depend on the joint effects of capacitances, resistances, and circuit sensitivities.

\section{H-IDET Systems Design}
\subsection{Near-filed Focus for Single EU}	
In the traditional near-field channel, the radius of the transmitting aperture is far less than the distance between the transmitter and the receiver. By applying Taylor series expansion $ \sqrt{1+x}=1+x/2+\mathcal{O}(x^2) $ in the traditional near-field channel \cite{1137900}, the near-field focusing is realized in the Fresnel region which ranges from  the radius of $ \frac{1}{2}\sqrt{\frac{D^3}{\lambda}} $ to $ 2\frac{D^2}{\lambda} $ \cite{10558818}. Different from the traditional channel, the Fresnel region in the EM channel is extended below the lower bound $\frac{1}{2}\sqrt{\frac{D^3}{\lambda}} $, where the EUs are more likely to be located. In order to highlight the difference between the traditional channel and the EM channel as well as analyze the near-filed focusing effect in the H-IDET system, we firstly investigate the maximum energy harvesting amount in the near-field region in this sub-section. In other words, we aim to maximize the actual harvested energy after the non-linear EH model, which is the same as maximizing the input energy harvesting power at the receiver, since $ \Xi(P_l) $ in Eq. (\ref{eq:8}) monotonically increases with  respect to $ P_l $.


Since the transmit power is limited in the IDET system, there is a constraint for the beamforming matrices \cite{10158997}: 
	\begin{equation}
		\sum^{K+L}_{j=1} \int_{\mathcal{S}_{\mathrm{ T}}} \left\|  \bm{\theta}_j(\mathbf{s})\right\| ^2 \mathrm {d}\mathbf{s} \leq P_t,
	\end{equation}
where $ P_t $ is the maximum transmit power of H-IDET, which is measured in the current form of $ \mathrm{A}^2 $ (or $ \mathrm{(mA)}^2 $). 

Consider the single EU is located in the radiation near-field, our goal is to obtain the optimal beamforming of the transmitter and polarization combining of the receiver, in order to maximize the harvested RF energy subject to the transmit power constraint, while the problem can be formulated as
\begin{align}
	\text{ (P1):}  & \max  \limits_{\bm{\theta}(\mathbf{s}),\bm{\psi}} \, \mathrm{P_{EH}}=\frac{A_{\text{R}}\cos(\phi) }{2Z}\left|\bm{\psi}^H  \int_{\mathcal{S}_{\mathrm{T}}} \mathbf{G}(\mathbf{s})\bm{\theta}(\mathbf{s})\mathrm{d}\mathbf{s} \right| ^2, \label{eq:object0} \\
	\text{s. t.} \  & \int_{\mathcal{S}_{\mathrm{ T}}}\left\| \bm{\theta}(\mathbf{s})\right\| ^2 \mathrm {d}\mathbf{s}\leq P
	_t,\tag{\ref{eq:object0}a} \label{eq:object0-a} \\
			&\left\| \bm{\psi} \right\|=1 ,  \tag{\ref{eq:object0}b} \label{eq:object0-b}
\end{align}  
where $ \bm{\psi} \in \mathbb{C}^{3\times 1} $ is the polarization combining vector at EU. (\ref{eq:object0-b}) represents that there is no additional power supplement at the receiver.

According to the Cauchy-Schwarz inequality : 
\begin{align}
	 \left|\bm{\psi}^H \!\! \int_{\mathcal{S}_{\mathrm{T}}}\!\! \mathbf{G}(\mathbf{s})\bm{\theta}(\mathbf{s})\mathrm{d}\mathbf{s} \right| ^2\!\!\leq \left| \int_{\mathcal{S}_{\mathrm{T}}}\!\! \left\| \bm{\theta}(\mathbf{s})\right\| ^2\mathrm{d}\mathbf{s}\right| \left|    \int_{\mathcal{S}_{\mathrm{T}}}\!\! \left\| \bm{\psi}^H\mathbf{G}(\mathbf{s})\right\|^2 \mathrm{d}\mathbf{s} \right|,
\end{align}
 when $ \bm{\psi} $ is fixed, $ \mathrm{P_{EH}} $ is maximized when 
\begin{equation}\label{eq:sigle-theta}
	\bm{\theta}(\mathbf{s})=\mu\mathbf{G}^H(\mathbf{s})\bm{\psi}
\end{equation}
where $ \mu =\sqrt{\frac{P_t}{\int_{\mathcal{S}_{\mathrm{T}}} \left\| \mathbf{G}^H(\mathbf{s'})\bm{\psi}\right\|^2\mathrm{d}\mathbf{s'}}}$ is the normalizing factor to satisfy the power constraint. By substituting (\ref{eq:sigle-theta}) into (\ref{eq:object0}) we have $ \mathrm{P_{EH}}=\frac{A_{\text{R}}\cos(\phi)P_t }{Z}\left\|\bm{\psi}  \int_{\mathcal{S}_{\mathrm{T}}} \mathbf{G}(\mathbf{s})\mathrm{d}\mathbf{s} \right\| ^2 $. Then, $ \mathrm{P_{EH}} $ is maximized when we have
\begin{equation}\label{eq:sigle-psi}
	\bm{\psi}=\frac{\xi_{max}\left\lbrace \int_{\mathcal{S}_{\mathrm{T}}}  \mathbf{G}(\mathbf{s})\mathbf{G}^H(\mathbf{s})\mathrm{d}\mathbf{s}\right\rbrace}{\left\|\xi_{max}\left\lbrace \int_{\mathcal{S}_{\mathrm{T}}}  \mathbf{G}(\mathbf{s})\mathbf{G}^H(\mathbf{s})\mathrm{d}\mathbf{s}\right\rbrace \right\|}
\end{equation}
where $ \bm{\xi}_{\mathrm{max}}\left\lbrace \cdot \right\rbrace  $ denotes the eigenvector corresponding to the maximum eigenvalue of the argument.

In Section III-A, the obtained closed-form beamforming can achieve single-point near-field focusing for WET. In Section III-B, the proposed BCD-based beamforming scheme will achieve multi-point near-field focusing simultaneously for the H-IDET system.

\subsection{H-IDET Design for Multi-user System}
\subsubsection{Problem Formulation} In a multi-user H-IDET system, we aim to obtain the optimal H-IDET beamformer and combiner, in order to maximize the sum-rate of the DUs by satisfying the EUs' minimum energy harvesting requirements. Therefore, the problem can be formulated as
\begin{align}
	\text{ (P2):}  & \max  \limits_{\boldsymbol{\Theta(\mathbf{s})}, \boldsymbol{\Psi}} \, R_{\mathrm{sum}} = \sum_{k=1}^{K}R_k, \label{eq:object1} \\
	\text{s. t.} \  & \sum_{k=1}^{K+L}\int_{\mathcal{S}_{\mathrm{ T}}}\left\| \bm{\theta}_k(\mathbf{s})\right\| ^2 \mathrm {d}\mathbf{s}\leq P
	_t,\tag{\ref{eq:object1}a} \label{eq:object1-a} \\
	&   \Xi(P_l) \geq  P_0, l \in \mathcal{E},  \tag{\ref{eq:object1}b} \label{eq:object1-b}\\
	&\left\| \bm{\psi}_l \right\|=1 , l \in \mathcal{E}, \tag{\ref{eq:object1}c} \label{eq:object1-c}
\end{align}
where $ \boldsymbol{\Theta(\mathbf{s})} $ is defined as $ \boldsymbol{\Theta(\mathbf{s})}:=\left\lbrace \bm{\theta}_k(\mathbf{s}) \right\rbrace_{k=1}^{K+L}  $, $\boldsymbol{\Psi} $ is defined as  $\boldsymbol{\Psi}:=\left\lbrace \boldsymbol{\psi}_k \right\rbrace_{k=1}^{K+L}$  and $ P_0 $ is the minimum required harvested power of the EUs. (\ref{eq:object1-c}) represents that there is no additional power supplement for EUs. Note that there is no constraints for the DUs' combiner $ \bm{\psi}_k, k\in\mathcal{D}$, since its norm value does not affect the achievable data rate according to Eq. (\ref{eq:sinr}). It can be observed that (P2) is a non-convex functional programming due to the following reasons. Firstly, the continuous beamform functions $ \boldsymbol{\Theta(\mathbf{s})} $ within integrals exists in both the objective and the constraint, so that the partial derivatives of the objective function with respect to $ \boldsymbol{\Theta(\mathbf{s})} $ are difficult to obtain. Secondly, the objective function is intractable and the optimization variables are coupled with each other.

\subsubsection{Problem Transformation}
To solve the non-convex functional programming (P2), we reformulate the problem by utilizing the equivalence between the achievable data rate and the MSE \cite{1223549}, since a lower MSE always results in a higher data rate. Denoting the decoded symbol as $ \hat{x_k}=\bm{\psi}_k^H \mathbf{y}_k $, the MSE at the $ k $-th DU is given by
\begin{align} \label{eq:11}
	{M_{k}} =&\, {\mathbb E_{{\mathbf{x}}, {\mathbf{n}}}}\left \{{ {\left |{ {\hat x_{k} - {x_{k}}} }\right |}^{2} }\right \} \nonumber \\ = &\,{\left |{ { \boldsymbol {\psi }}_{k}^H {\int _{{S_{\mathrm{T}}}} {{{\mathbf{G}}_{k}}({\mathbf{s}})} { \boldsymbol \theta _{k}}\left ({{\mathbf{s}} }\right){\mathrm{d}}{\mathbf{s}}}-1 }\right |^{2}} \nonumber \\+& \sum \limits _{j = 1,j \ne k}^{K+L} {{{\left |{ { \boldsymbol {\psi }}_{k}^{H}{\int _{{S_{\mathrm{T}}}} {{{\mathbf{G}}_{k}}({\mathbf{s}}){{ \boldsymbol {\theta }}_{j}}\left ({{\mathbf{s}} }\right)} {\mathrm{d}}{\mathbf{s}}} }\right |}^{2}}} \!+\! {\sigma ^{2}}{\left \|{ {{{ \boldsymbol {\psi }}_{k}}} }\right \|^{2}}. 
\end{align}
\begin{lemma}
	By introducing auxiliary variables $ \boldsymbol{\rho}:=\left\lbrace \rho_k \in \mathbb{R}^+ \right\rbrace_{k=1}^{K}$ and the combining verctors $ \bm{\Psi}:=\left\lbrace \bm{\psi}_k\right\rbrace_{k=1}^{K+L}  $ for all users, the problem (P2) can be reformulated as
	\begin{align} 
		\text{ (P3):} &\min \limits _{ \boldsymbol{\rho}, { \boldsymbol \Psi }, \boldsymbol{\Theta(\mathbf{s})}}     \sum \limits _{k = 1}^{K} {\rho _{k}{M_{k}}}-\sum \limits _{k = 1}^{K} {\log {\rho _{k}}}\label{eq:object2} \\ 
		\text{s. t.} &\sum \limits _{k = 1}^{K+L} {\int _{{\cal S_{\mathrm{T}}}} {{{\left \|{ {{{ \boldsymbol {\theta }}_{k}}\left ({{\mathbf{s}} }\right)} }\right \|}^{2}}{\mathrm{d}}{\mathbf{s}}} } \le {P_t}, \tag{\ref{eq:object2}a} \label{eq:object2-a}\\
		& \sum^{K+L}_{j=1}\left\| \bm{\psi}_l^H \int_{\mathcal{S}_{\mathrm{T}}}  \mathbf{G}_l(\mathbf{s})\bm{\theta}_j(\mathbf{s})\mathrm{d}\mathbf{s} \right\| ^2 \geq P_0^{'}, l \in \mathcal{E},  \tag{\ref{eq:object2}b} \label{eq:object2-b}\\
		&\left\| \bm{\psi}_l \right\|=1 , l \in \mathcal{E}, \tag{\ref{eq:object2}c} \label{eq:object2-c}
	\end{align}
	where $ P_0^{'}= \frac{A_{\text{R}}\cos(\phi)}{Z}\Xi^{-1}(P_0) $ and $ \Xi^{-1} $ is inverse function of $ \Xi $ in Eq. (\ref{eq:8}).  
\end{lemma}

\begin{IEEEproof}
	We aim to prove that (P3) can be transformed into (P2), which indicates that these two problems are equivalent. According to \cite{Tse_Viswanath_2005}, The optimal minimum MSE (MMSE) combining verctor of the DUs is given by
	\begin{align}\label{eq:13}
		\bm{\psi}_k&=\left( \sum_{j=1}^{K+L}\int_{\mathcal{S}_{\mathrm{T}}}  \mathbf{G}_k(\mathbf{s})\bm{\theta}_j(\mathbf{s})\mathrm{d}\mathbf{s}\left( \int_{\mathcal{S}_{\mathrm{T}}}  \mathbf{G}_k(\mathbf{s})\bm{\theta}_j(\mathbf{s})\mathrm{d}\mathbf{s}\right) ^H +\sigma^2 \mathbf{I}_3	 \right)^{-1}\nonumber\\
		&\times\int_{\mathcal{S}_{\mathrm{T}}}  \mathbf{G}_k(\mathbf{s})\bm{\theta}_k(\mathbf{s})\mathrm{d}\mathbf{s} ,
	\end{align}	
	 which is also the optimum $ \bm{\psi}_k $ for (P3). By substituting (\ref{eq:13}) into (\ref{eq:11}), the corresponding MSE is obtained as 
	 	\begin{align}\label{eq:14}
	 	M_k^{\mathrm{min}}&= - \left( \int_{\mathcal{S}_{\mathrm{T}}}  \mathbf{G}_k(\mathbf{s})\bm{\theta}_j(\mathbf{s})\mathrm{d}\mathbf{s}\right) ^H \nonumber \\ &\times \left( \sum_{j=1}^{K+L}\int_{\mathcal{S}_{\mathrm{T}}}  \mathbf{G}_k(\mathbf{s})\bm{\theta}_j(\mathbf{s})\mathrm{d}\mathbf{s}\left( \int_{\mathcal{S}_{\mathrm{T}}}  \mathbf{G}_k(\mathbf{s})\bm{\theta}_j(\mathbf{s})\mathrm{d}\mathbf{s}\right) ^H \!\!\!+\sigma^2 \mathbf{I}_3	 \right)^{-1} \nonumber \\ &\times \int_{\mathcal{S}_{\mathrm{T}}}  \mathbf{G}_k(\mathbf{s})\bm{\theta}_j(\mathbf{s})\mathrm{d}\mathbf{s}+1.
	 \end{align}
	 The optimal combining verctor for EUs is given by
	 \begin{align}\label{eq:15}
	 	\bm{\psi}_l&=\mathrm{arg}\max\sum^{K+L}_{j=1}\left\| \bm{\psi}_l \int_{\mathcal{S}_{\mathrm{T}}}  \mathbf{G}_l(\mathbf{s})\bm{\theta}_j(\mathbf{s})\mathrm{d}\mathbf{s} \right\| ^2 \nonumber \\
	 	&=\dfrac{\bm{\xi_{\mathrm{max}}}\left\lbrace \sum\limits_{j=1}^{K+L}\int_{\mathcal{S}_{\mathrm{T}}}  \mathbf{G}_l(\mathbf{s})\bm{\theta}_j(\mathbf{s})\mathrm{d}\mathbf{s}\left( \int_{\mathcal{S}_{\mathrm{T}}}  \mathbf{G}_l(\mathbf{s})\bm{\theta}_j(\mathbf{s})\mathrm{d}\mathbf{s}\right) ^H \right\rbrace }{\left\| \bm{\xi_{\mathrm{max}}}\left\lbrace \sum \limits _{j=1}^{K+L}\int_{\mathcal{S}_{\mathrm{T}}}  \mathbf{G}_l(\mathbf{s})\bm{\theta}_j(\mathbf{s})\mathrm{d}\mathbf{s}\left( \int_{\mathcal{S}_{\mathrm{T}}}  \mathbf{G}_l(\mathbf{s})\bm{\theta}_j(\mathbf{s})\mathrm{d}\mathbf{s}\right) ^H \right\rbrace \right\| },
	 \end{align}
	 According to the first order optimality condition for $ \rho_k $ in (P3), the optimal $ \rho_k^* $ is given by 
	 \begin{equation}\label{eq:16}
	 	\rho_k^* =(M_k^{\mathrm{min}})^{-1}.
	 \end{equation}
     By substituting (\ref{eq:16}), (\ref{eq:14}) and (\ref{eq:15}) into (P3) and utilizing $\mathrm{det}| \mathbf{I}+\mathbf{AB}|=\mathrm{det}|\mathbf{I}+\mathbf{BA} | $ and  the matrix inversion lemma, i.e.,  $  (\mathbf{A}+\mathbf{BCD})^{-1}=\mathbf{A}^{-1}-\mathbf{A}^{-1}\mathbf{B}(\mathbf{C}^{-1}+\mathbf{D}\mathbf{A}^{-1}\mathbf{B})^{-1}\mathbf{D}\mathbf{A}^{-1} $, the object of (P3) is derived as
     \begin{equation}
     	K-\sum_{k=1}^{K}\log_2\det\left| \mathbf {I_3}+ \int_{\mathcal{S}_{\mathrm{ T}}} \mathbf{G}_k(\mathbf{s}) \bm{\theta}_k(\mathbf{s})\mathrm {d}\mathbf{s} \left( \int_{\mathcal{S}_{\mathrm{ T}}} \mathbf{G}_k(\mathbf{s'}) \bm{\theta}_k(\mathbf{s'})\mathrm {d}\mathbf{s'} \right) ^H \mathbf{J}_k^{-1}  \right| 
     \end{equation}
    where $ \mathbf{J}_k=\sum^{K+L}_{j=1,j \neq k} \int_{\mathcal{S}_{\mathrm{ T}}} \mathbf{G}_k(\mathbf{s})\bm{\theta}_j(\mathbf{s})\mathrm {d}\mathbf{s}\left( \int_{\mathcal{S}_{\mathrm{ T}}} \mathbf{G}_k(\mathbf{s'}) \bm{\theta}_k(\mathbf{s'})\mathrm {d}\mathbf{s'} \right) ^H+\sigma^2 \mathbf {I_3} $. 
     Then, by removing the constant term and introducing the polarization combinner $ \bm{\psi} $, (P3) is transformed into (P2), which completes the proof.
\end{IEEEproof}

\subsubsection{Holographic Beamforming Design}
The block coordinate descent (BCD) algorithm is conceived to solve problem (P3), where the variables are separated and optimized iteratively by fixing the other one as the constant. By introducing the auxiliary variables $ \rho $ in (P3) and based on BCD algorithm, the subproblem of (P3) without $ \log $ function can be optimized. Specifically, given the beamforming set $\boldsymbol{\Theta}(\mathbf{s})$, $ \bm{\rho} $ and $ \bm{\Psi} $ are already obtained by (\ref{eq:13}), (\ref{eq:15}) and (\ref{eq:16}) respectively in the proof of \textbf{Lemma 1}. In the context, we aim to optimize $\boldsymbol{\Theta}(\mathbf{s})$ by fixing $\bm{\rho}$ and $\bm{\Psi}$. By substituting the MSE $ M_k $ in (\ref{eq:11}) into (\ref{eq:object2}) and discarding the constant terms, we obtain
	{\small 
		\begin{align}\label{eq:24}
			&\sum \limits _{k = 1}^{K} \rho_{k}\left( \,{\left |{ { \boldsymbol {\psi }}_{k}^H {\int _{{S_{\mathrm{T}}}} {{{\mathbf{G}}_{k}}({\mathbf{s}})} { \boldsymbol \theta _{k}}\left ({{\mathbf{s}} }\right){\mathrm{d}}{\mathbf{s}}}-1 }\right |^{2}} \nonumber \!\!+\!\! \sum \limits _{j = 1,j \ne k}^{K+L} {{{\left |{ { \boldsymbol {\psi }}_{k}^{H}{\int _{{S_{\mathrm{T}}}} {{{\mathbf{G}}_{k}}({\mathbf{s}}){{ \boldsymbol {\theta }}_{j}}\left ({{\mathbf{s}} }\right)} {\mathrm{d}}{\mathbf{s}}} }\right |}^{2}}}\right) 
		\end{align}
		\begin{align}
			=& \sum \limits _{k = 1}^{K} \rho _{k}\left(   \sum_{j=1}^{K+L}\left|\bm{\psi}_k^H \int_{\mathcal{S}_{\mathrm{T}}} \!\! \mathbf{G}_k(\mathbf{s})\bm{\theta}_j(\mathbf{s})\mathrm{d}\mathbf{s}\right| ^2 \!\!\!-\! 2\mathfrak{R}\left\lbrace \bm{\psi}_k^H \int_{\mathcal{S}_{\mathrm{T}}} \!\! \mathbf{G}_k(\mathbf{s})\bm{\theta}_k(\mathbf{s})\mathrm{d}\mathbf{s}\right\rbrace  \right)  + \sum \limits _{k = 1}^{K} \rho_{k}
\end{align}
where the real term $ 2\mathfrak{R}\left\lbrace \bm{\psi}_k^H \int_{\mathcal{S}_{\mathrm{T}}} \!\! \mathbf{G}_k(\mathbf{s})\bm{\theta}_k(\mathbf{s})\mathrm{d}\mathbf{s}\right\rbrace  $ is obtained form $  { \boldsymbol {\psi }}_{k}^H {\int _{{S_{\mathrm{T}}}} {{{\mathbf{G}}_{k}}({\mathbf{s}})} { \boldsymbol \theta _{k}}\left ({{\mathbf{s}} }\right){\mathrm{d}}{\mathbf{s}}}+ \left(  {\int _{{S_{\mathrm{T}}}} {{{\mathbf{G}}_{k}}({\mathbf{s}})} { \boldsymbol \theta _{k}}\left({{\mathbf{s}} }\right){\mathrm{d}}{\mathbf{s}}} \right) ^H { \boldsymbol {\psi }}_{k} $. By discarding the constant terms in (\ref{eq:24}), the problem (P3) is then re-formulated as
	\begin{align} 
	\text{ (P4):} &\min \limits _{  \boldsymbol{\Theta(\mathbf{s})}}     \sum \limits _{k = 1}^{K} \rho _{k}\Bigg(  \sum_{j=1}^{K+L}\left|\bm{\psi}_k^H \int_{\mathcal{S}_{\mathrm{T}}}  \mathbf{G}_k(\mathbf{s})\bm{\theta}_j(\mathbf{s})\mathrm{d}\mathbf{s}\right| ^2 \nonumber \\&\ \ \ \ \ -2\mathfrak{R}\left\lbrace \bm{\psi}_k^H \int_{\mathcal{S}_{\mathrm{T}}}  \mathbf{G}_k(\mathbf{s})\bm{\theta}_k(\mathbf{s})\mathrm{d}\mathbf{s}\right\rbrace  \Bigg) \label{eq:object3} \\ 
	\text{s. t.} &\sum \limits _{k = 1}^{K+L} {\int _{{\cal S_{\mathrm{T}}}} {{{\left \|{ {{{ \boldsymbol {\theta }}_{k}}\left ({{\mathbf{s}} }\right)} }\right \|}^{2}}{\mathrm{d}}{\mathbf{s}}} } \le {P_t}, \tag{\ref{eq:object3}a} \label{eq:object3-a}\\
	& \sum^{K+L}_{j=1}\left\| \bm{\psi}_l^H \int_{\mathcal{S}_{\mathrm{T}}}  \mathbf{G}_l(\mathbf{s})\bm{\theta}_j(\mathbf{s})\mathrm{d}\mathbf{s} \right\| ^2 \geq P_0^{'}, l \in \mathcal{E},  \tag{\ref{eq:object3}b} \label{eq:object3-b}
\end{align}}

Then, the Fourier analysis is relied upon for solving (P4), which is widely used in EM theory based holographic systems \cite{1386525,4685903,9724113,9906802,10158997,9139337}. With the aid of the mapping between wavenumber domains and spatial domains by the Fourier transform, the performance analysis and communication scheme designs can be moved into wavenumber domains. Typically, the beamforming function for each user can be expanded by finite-item Fourier series approximation as \cite{10158997}:
\begin{align} \label{eq:23}
	{ \boldsymbol \theta }_{k}({\mathbf{s}})\approx&\sum \limits _{\mathbf{n}}^{\mathbf{N}} {{{\mathbf{w}}_{k, {\mathbf{n}}}}{\Upsilon _{\mathbf{n}}}\left ({{\mathbf{s}} }\right)}, {\mathbf{s}} \in {\mathcal{ S}}_{\mathrm{ T}}, \forall k,j\in \{1,\cdots,K+L\},
\end{align} 
where $ {\Upsilon _{\mathbf{n}}}\left({{\mathbf{s}} }\right)=\frac {1}{\sqrt {A_{\mathrm{ T}}}}{e^{ 2\pi {\mathrm{ j}} \left ({{\frac {n_{x}}{L_{x}}\left ({s_{x}-\frac {L_{x}}{2}}\right) + \frac {n_{y}}{L_{y}}{\left ({s_{y}-\frac {L_{y}}{2}}\right)} + \frac {n_{z}}{L_{z}}{\left ({s_{z}-\frac {L_{z}}{2}}\right)}} }\right)}} \in \mathbb{C}$ \footnote{The orthogonal basis has a similar form mathematically to the DFT beam. However, the beamforming $ \bm{\theta}(\mathbf{s}) $ is expanded for each point on the continuous aperture and the number of bases $ N_F $ is much smaller than the number of points. Moreover, the basis functions only provide functional degrees of freedom for the further optimization of beamforming functions $ \bm{\theta}(\mathbf{s}) $, which have no physical significance.} is the Fourier orthogonal base function, and $ {{\mathbf{w}}_{k, {\mathbf{n}}}} = \frac {1}{\sqrt {A_{\mathrm{ T}}}}\int _{{\mathbf{s}} \in {\mathcal{ S}}_{\mathrm{ T}}} { \boldsymbol \theta _{k}({\mathbf{s}}){\Upsilon ^{\ast}_{\mathbf{n}}}\left ({{\mathbf{s}} }\right){\mathrm{ d}}{\mathbf{s}}} \in \mathbb{C}^{3\times 1}$ is the projection length in the wavenumber domain. For the simplicity, we define $ \mathbf{n}:=(n_x,n_y,n_z) \in \mathbb{Z}^{3\times 1} $, $ \mathbf{N}:=(N_x,N_y,N_z) \in \mathbb{Z}^{3\times 1} $, $ \sum  _{\mathbf{n}}^{\mathbf{N}}:=\sum  _{n_x=-N_x}^{N_x}\sum  _{n_y=-N_y}^{N_y}\sum   _{n_z=-N_z}^{N_z} $. The total number of Fourier bases is $ N_F=(2N_x+1)(2N_y+1)(2N_z+1) $. Besides, $ L_x $, $ L_y $ and $ L_z $ denote the maximum projection lengths of $ \mathcal{ S}_{\mathrm{ T}} $ on 3D cartesian coordinate system, respectively. Then, in the electric field and power constraint, Fourier transform can be applied as \cite{10158997}:
\begin{align}
	\!\!\!\int _{{\cal S_{\mathrm{T}}}}\! {{{\mathbf{G}}_{k}}({\mathbf{s}}){{ \boldsymbol {\theta }}_{j}}\left ({{\mathbf{s}} }\right){\mathrm{ d}}{\mathbf{s}}} &\approx \sum \limits _{\mathbf{n}}^{\mathbf{N}} {{{\boldsymbol{\Omega }}_{k, {\mathbf{n}}}}{{\mathbf{w}}_{j, {\mathbf{n}}}}}, \forall k,j\in \{1,\cdots,K+L\},  \label{eq:18}
\end{align}
\begin{align}
	\int _{{\cal S_{\mathrm{T}}}}\!\! {{{\left \|{ {{{ \boldsymbol {\theta }}_{k}}\left ({{\mathbf{s}} }\right)} }\right \|}^{2}}{\mathrm{d}}{\mathbf{s}}} 
	&\approx \sum \limits _{\mathbf{n}}^{\mathbf{N}} {{{\left \|{ {{{\mathbf{w}}_{k, {\mathbf{n}}}}} }\right \|}^{2}}}, \forall k\in \{1,\cdots,K+L\}, \label{eq:19}
\end{align}
where 
\begin{align} 
	{{\boldsymbol{\Omega }}_{k, {\mathbf{n}}}} =& \int _{{\cal S_{\mathrm{T}}}} {{{\mathbf{G}}_{k}}({\mathbf{s}}){\Upsilon _{\mathbf{n}}}\left ({{\mathbf{s}} }\right){\mathrm{d}}{\mathbf{s}}}, 
\end{align}\footnote{The conventional DFT beam can only be used in the far field because the channel response vector linear with respect to the antenna index is used. However, the channel characterization in wavenumber space is applicable in both the far-field and near-field.}
is the Fourier transform of the EM channel $ {{\mathbf{G}}_{k}}({\mathbf{s}}) $. 

\begin{remark}
	Thanks to the inherent physical properties of electromagnetic channel function $ \mathbf{G}_k(\mathbf{s}) $, some recent works have revealed that, the value of channel gain $ \left\| \boldsymbol{\Omega }_{k, {\mathbf{n}}} \right\|^2  $ in the wavenumber domain, is high in the low-wavenumber band and is negligible in the high-wavenumber band in most cases \cite{9906802,10158997}. Therefore, the functions $ \left\lbrace \bm{\theta}_k(\mathbf{s}) \right\rbrace_{k=1}^{K+L} $ and $ \left\lbrace \mathbf{G}_k(\mathbf{s}) \right\rbrace_{k=1}^{K+L}$ can be approximated by finite low-wavenumber and high-power items. Specifically, when $ N_x = \lceil\frac{L_x}{\lambda}\rceil$, $ N_y=\lceil\frac{L_y}{\lambda}\rceil $ and $ N_z=\lceil\frac{L_z}{\lambda}\rceil $, i.e., the number of finite terms $ N_F $ is not less than $ (2\lceil\frac{L_x}{\lambda}\rceil+1)(2\lceil\frac{L_y}{\lambda}\rceil+1)(2\lceil\frac{L_z}{\lambda}\rceil+1) $, there are sufficient functional degrees of freedom for approximation \cite{9906802,10158997}. Thus, the performance loss can be artificially controlled by presetting an acceptable number of finite terms $ N_F $.
\end{remark}

 
Based on (\ref{eq:18}) and (\ref{eq:19}), the functional programming (P4) can be then reformulated as
\begin{align}
	\text{ (P5):}  & \ { R_{\mathrm{eq}}=  \min \limits_{\boldsymbol{w}}} \sum_{k=1}^{K}\rho_k(\sum_{j=1}^{K+L} |\mathbf{h}_k^H \mathbf{w}_j|^2-2\mathfrak{R}(\mathbf{h}_k^H \mathbf{w}_k)), \label{eq:object4} \\
	\text{s. t.} \  & \sum_{k=1}^{K+L}\|\mathbf{w}_k\|\leq P
	_t, \tag{\ref{eq:object4}a} \label{eq:object4-a} \\
	& \sum_{j=1}^{K+L}\left\| \mathbf{h}_l^H \mathbf{w}_j\right\| ^2 \geq P_0^{'} , l \in \mathcal{E}, \tag{\ref{eq:object4}b} \label{eq:object4-b}
\end{align}
where $ \mathbf{h}_k $ and $ \mathbf{w}_k $ are the vectorized sets of $ \mathbf{h}_{k,\mathbf{n}} $ and $ \mathbf{w}_{k,\mathbf{n}} $ for all $ \mathbf{n} \in \{ \{ -N_x,\cdots,N_x \} ,  \{ -N_y,\cdots,N_y \}, \{ -N_z,\cdots,N_z \} \} $ respectively, i.e., $  \mathbf{h}_k =\left[\mathbf{h}_{k,\mathbf{1}} ;\dots; \mathbf{h}_{k,\mathbf{N}} \right]  $ and $ \mathbf{w}_{k} =\left[\mathbf{w}_{k,\mathbf{1}} ;\dots; \mathbf{w}_{k,\mathbf{N}} \right] $. $ \mathbf{h}_{k,\mathbf{n}} :=\boldsymbol{\Omega }_{k, {\mathbf{n}}}^H \bm{\psi}_k $ and $ \boldsymbol{w}:=\{ \mathbf{w}_k \}_{k=1}^{K+L} $.

\begin{remark}
	By using the Fourier transform with finite number of bases, the functional programming (P4) is transformed into the classic quadratically constrained quadratic programming (QCQP) (P5). Therefore, the classic schemes for solving the QCQP can be adopted to solve it. However, there is a performance gap between (P4) and (P5), which is dependent with the number $ N_F $ of orthogonal bases. Since $ N_F $ represents the accuracy of approximation in Eq. (\ref{eq:23}), the gap can be narrowed as $ N_F $ increases, which is also shown in Fig. (\ref{fig_energy}). Besides, a lager $ N_F $ also results in a higher computational complexity, which is showed in the complexity analysis in Section III-(5).
	
	Furthermore, different from the physical channel in conventional problems, $ \mathbf{h}_k $ in (P5) is the equivalent counterpart of the physical electromagnetic channel after the Fourier transform and polarization combining. Moreover, the variable $ \boldsymbol{w} $ in (P5) is the projection length of $ \bm{\theta}(\mathbf{s}) $ on different orthogonal bases, while the variables in conventional problems are usually matrix of digital precoder and phase shifters.
	
	Besides, conventional MIMO systems consist of discrete antenna arrays with half wavelength spacing and is fed in parallel. Holographic MIMO systems use metasurface antennas and break through the limit of half wavelength spacing. The near-continuous aperture enables powerful electromagnetic manipulation capabilities, which can further approach the performance limit. The practical hardware of holographic MIMO mostly uses series feeding, which has higher energy efficiency.
\end{remark}

The problem (P5) is a QCQP. However, the energy harvesting constraints (\ref{eq:object4-b}) are still non-convex, so that the successive convex approximation (SCA) method is then employed for obtaining the solution. Given the beamfroming vector $ \mathbf{\bar{w}}_j $, the first-order Taylor expansion is expressed as
\begin{align}\label{eq:29}
	\left\| \mathbf{h}_l^H \mathbf{w}_j\right\| ^2 \geq  2\mathfrak{R}(\mathbf{\bar{w}}_j^H\mathbf{h}_l \mathbf{h}_l^H \mathbf{w}_j)-\mathbf{\bar{w}}_j^H\mathbf{h}_l \mathbf{h}_l^H \mathbf{\bar{w}}_j.
\end{align}
Further, (P5) can be transformed into a series of convex problems with given $ \mathbf{\bar{w}}_j $, which is reformulated as 
\begin{align}
		\text{ (P6):}  &  \ { \min \limits_{\boldsymbol{w}}} \sum_{k=1}^{K}\rho_k(\sum_{j=1}^{K+L} |\mathbf{h}_k^H \mathbf{w}_j|^2-2\mathfrak{R}(\mathbf{h}_k^H \mathbf{w}_k)), \label{eq:object5} \\
		\text{s. t.} \  & \sum_{k=1}^{K+L}\|\mathbf{w}_k\|\leq P
		_t, \tag{\ref{eq:object5}a} \label{eq:object5-a} \\
		&\sum_{j=1}^{K+L}(2\mathfrak{R}(\mathbf{\bar{w}}_j^H\mathbf{h}_l \mathbf{h}_l^H \mathbf{w}_j)-\mathbf{\bar{w}}_j^H\mathbf{h}_l \mathbf{h}_l^H \mathbf{\bar{w}}_j) \geq P_0^{'} , l \in \mathcal{E}.\tag{\ref{eq:object5}b} \label{eq:object5-b}
\end{align}
Note that (P6) is a convex optimization problem and can be solved by standard tools, such as the CVX. Finally, by substituting $ \boldsymbol{w} $ into (\ref{eq:23}), the H-IDET beamformer $ \boldsymbol{\Theta(\mathbf{s})} $ is obtained. The details of the algorithm is summarized in Algorithm I.

\subsubsection{Variables Initialization}
However, when the maximum transmit power $ P_t $ is too small, it is hard to satisfy the EUs' WET requrements, so that there is no feasible solutions for (P5) and (P6). Moreover, the initial value of $ \boldsymbol{\bar{w}} :=\{ \mathbf{\bar{w}}_k \}_{k=1}^{K+L} $ in the iteration has a great influence on the solution of (P6). It is necessary to select an optimal initial value of $ \boldsymbol{\bar{w}}^{*} $ for the EUs to make (P4) as solvable as possible, while it can be obtained by maximizing the minimum harvested energy of the EUs, which is formulated as 
\begin{align}
	\text{ (P7):}  &  \ { \boldsymbol{\bar{w}}^{*}=\mathop{\arg\max} \limits_{\boldsymbol{w} }}\min\limits_{l} \sum_{j=1}^{K+L}\left\| \mathbf{h}_l^H \mathbf{w}_j\right\| ^2, l \in \mathcal{E} , \label{eq:object6} \\
	\text{s. t.} \  & \sum_{k=1}^{K+L}\left\| \mathbf{w}_k\right\| \leq P_t . \tag{\ref{eq:object6}a} \label{eq:object6-a}
\end{align}
Similarly, by using the first-order expansion in (\ref{eq:29}), (P7) can be transformed into a series of convex problems with given $ \mathbf{\bar{w}}_j $, which is reformulated as 
\begin{align}
	\text{ (P8):}  &  \  \mathop{\max}\limits_{\boldsymbol{w}} \gamma \label{eq:object7} \\ \text{s. t.} \ 
	&  \sum_{j=1}^{K+L}(2\mathfrak{R}(\mathbf{\bar{w}}_j^H\mathbf{h}_l \mathbf{h}_l^H \mathbf{w}_j)-\mathbf{\bar{w}}_j^H\mathbf{h}_l \mathbf{h}_l^H \mathbf{\bar{w}}_j) \geq \gamma , l \in \mathcal{E}, \tag{\ref{eq:object7}a} \label{eq:object7-a} \\
	 & \sum_{k=1}^{K+L}\left\| \mathbf{w}_k\right\| \leq P_t . \tag{\ref{eq:object7}b} \label{eq:object7-b}
\end{align}
Then, (P8) is a convex optimization problem and can be solved by standard tools such as CVX. In addition to initializing $ \boldsymbol{\bar{w}}^{*} $ for obtaining the holographic beamforming, the initialization of global value of  $ \bm{\bar{\Psi}} $ and $ \boldsymbol{\bar{\Theta}} $ for BCD iterations also needs to be further considered to improve the robustness of our algorithm. The initial values of $ \boldsymbol{\bar{\Psi}} $ and $ \boldsymbol{\bar{\Theta}} $ are obtained similarly by maximizing the minimum harvested  energy of the EUs, which is formulated as 
\begin{align}
	\text{ (P9):}  & \max  \limits_{	\mathbf{j(s)},\bm{\Psi}} \min  \limits_{l}\, \frac{A_{\text{R}}\cos(\phi_l) }{2Z}\left\|\bm{\psi_l}^H  \int_{\mathcal{S}_{\mathrm{T}}} \mathbf{G}_l(\mathbf{s})	\mathbf{j(s)}\mathrm{d}\mathbf{s} \right\| ^2, \label{eq:object9} 
\end{align}
\begin{align}
	\text{s. t.} \  & \int_{\mathcal{S}_{\mathrm{ T}}}\left\| \mathbf{j}(\mathbf{s})\right\| ^2 \mathrm {d}\mathbf{s}\leq P
	_t,\tag{\ref{eq:object9}a} \label{eq:object9-a} 
 \end{align}  
Then, we exploit the spatial orthogonality to obtain a low-complexity solution. 
\begin{lemma}
When the transmit antenna has infinite aperture, the channels from the transmitter to different users are orthogonal to each other \cite{10315138}, which can be expressed as 
	\begin{equation}
		\lim_{| \mathcal{S}_{\mathrm{ T}} | \rightarrow \infty } \frac{\left\|\int_{\mathcal{S}_{\mathrm{ T}}} \mathbf{G}_i(\mathbf{s})\mathbf{G}^H_j(\mathbf{s}) \mathrm {d}\mathbf{s} \right\| ^2}{| \mathcal{S}_{\mathrm{ T}} |}=0,\ \mathrm{for} \  i \neq j.
	\end{equation}
where $ \mathbf{G}_i $ and $ \mathbf{G}_j $ are the channels spanning from the transmitter to two different users, respectively.
\end{lemma}
\begin{proposition}
The asymptotically optimal solution for (P9) can be expressed as  
\begin{equation}\label{eq:35}
		\mathbf{j(s)}=\sum_{l=1}^{L}\beta_l\mathbf{G}^H_l(\mathbf{s})\bm{\psi_l},
\end{equation}
\begin{equation}\label{eq:36}
	\bm{\psi_l}=\frac{\xi_{max}\left\lbrace \int_{\mathcal{S}_{\mathrm{T}}}  \mathbf{G}_l(\mathbf{s})\mathbf{G}_l^H(\mathbf{s})\mathrm{d}\mathbf{s}\right\rbrace}{\left\|\xi_{max}\left\lbrace \int_{\mathcal{S}_{\mathrm{T}}}  \mathbf{G}_l(\mathbf{s})\mathbf{G}_l^H(\mathbf{s})\mathrm{d}\mathbf{s}\right\rbrace \right\|}.
\end{equation}
	where $ \beta_l=\sqrt{\dfrac{P_t}{\zeta_l^2\sum_{i=1}^{L}\frac{1}{\zeta_i}}} $ and $ \zeta_i=\left|\int_{\mathcal{S}_{\mathrm{ T}}}\bm{\psi}^H_i \mathbf{G}_i(\mathbf{s})\mathbf{G}^H_i\bm{\psi}_i(\mathbf{s}) \mathrm {d}\mathbf{s} \right|^2  $.
\end{proposition}
\begin{IEEEproof}
	According to {\it Lemma} 2, $ \{\mathbf{G}^H_l(\mathbf{s})\bm{\psi_l}| l \in \mathcal{E}\} $ are asymptotically orthogonal to one another. Then the optimal solution for (P9) can be expressed as 
	\begin{equation}
 	\mathbf{j(s)}=\sum_{l=1}^{L}\beta_l\mathbf{G}^H_l(\mathbf{s})\bm{\psi_l}+\sum_{i}\tau_i\mathbf{g}_i ,
	\end{equation}
 where $ \mathbf{g}_i $ are vectors orthogonal to $ \{\mathbf{G}^H_l(\mathbf{s})\bm{\psi_l}| l \in \mathcal{E}\} $.
	Further, the asymptotically harvested RF power by the $ l $-th EU can be expressed as 
	\begin{align}
		E_l&=\lim_{| \mathcal{S}_{\mathrm{ T}} | \rightarrow \infty }\frac{A_{\text{R}}\cos(\phi_l) }{2Z}\Bigg\|\bm{\psi_l}^H  \int_{\mathcal{S}_{\mathrm{T}}} \mathbf{G}_l(\mathbf{s})\Bigg(\beta_l\mathbf{G}^H_l(\mathbf{s})\bm{\psi_l}\nonumber\\&+ \sum_{j=1,j\neq l}^{L}\beta_j\mathbf{G}^H_j(\mathbf{s})\bm{\psi_j}+\sum_{i}\tau_i\mathbf{g}_i\Bigg) \mathrm{d}\mathbf{s} \Bigg\| ^2 \nonumber \\ 
		&=\frac{A_{\text{R}}\cos(\phi_l)\beta_l^2 }{2Z}\left\|  \int_{\mathcal{S}_{\mathrm{T}}}\bm{\psi_l}^H \mathbf{G}_l(\mathbf{s})\mathbf{G}^H_l(\mathbf{s})\bm{\psi_l} \mathrm{d}\mathbf{s} \right\| ^2.
	\end{align}
Therefore, the asymptotically optimal $ \bm{\psi_l} $ for the EU$_l $ can be obtained by $ \bm{\psi_l}=\arg\max E_l=\frac{\xi_{max}\left\lbrace \int_{\mathcal{S}_{\mathrm{T}}}  \mathbf{G}_l(\mathbf{s})\mathbf{G}_l^H(\mathbf{s})\mathrm{d}\mathbf{s}\right\rbrace}{\left\|\xi_{max}\left\lbrace \int_{\mathcal{S}_{\mathrm{T}}}  \mathbf{G}_l(\mathbf{s})\mathbf{G}_l^H(\mathbf{s})\mathrm{d}\mathbf{s}\right\rbrace \right\|}$. Since the second term of $ \mathbf{j(s)} $ have no contribution to any user, we set $ \tau_i=0 $. Moreover, the asymptotically optimal $ 	\mathbf{j(s)} $ is obtained when all EUs harvest the same energy. By considering the constraint of (\ref{eq:object9}a), we have the following equation
\begin{equation}
	\begin{cases}
\int_{\mathcal{S}_{\mathrm{ T}}}\left\| \sum_{l=1}^{L}\beta_l\mathbf{G}^H_l(\mathbf{s})\bm{\psi_l}\right\| ^2 \mathrm {d}\mathbf{s}= P_t, \\
E_i=E_j, \forall i,j \in \mathcal{E}.
	\end{cases}
\end{equation}
Then, $ \mathbf{j(s)}$ can be obtained by solving the equation, which is experssed as $ 	\mathbf{j(s)}=\sum_{l=1}^{L}\beta_l\mathbf{G}^H_l(\mathbf{s})\bm{\psi_l} $, where $ \beta_l=\sqrt{\dfrac{P_t}{\zeta_l^2\sum_{i=1}^{L}\frac{1}{\zeta_i}}} $ and $ \zeta_i=\left|\int_{\mathcal{S}_{\mathrm{ T}}}\bm{\psi}^H_i \mathbf{G}_i(\mathbf{s})\mathbf{G}^H_i\bm{\psi}_i(\mathbf{s}) \mathrm {d}\mathbf{s} \right|  $.
\end{IEEEproof}

Further $ \bm{\theta}_k(\mathbf{s}) $ can be initialized by $ \bm{\theta}_k(\mathbf{s})=m\sum_{l=1}^{L}\beta_l\mathbf{G}^H_l(\mathbf{s})\bm{\psi_l}, \forall k \in  \mathcal{U}$, where $ m $ is the scaling parameter to satisify the power constrains.

For clarity, the details of our algorithm for obtaining the optimal $ R_{\mathrm{sum}} $, $ \bm{\Psi}$ and $ \boldsymbol{\Theta}$ are summarized in Algorithm I.


\begin{algorithm}[!t]
	\linespread{1}
	\caption{The Holographic Beamforming Design for IDET Systems}
	\small
	\begin{algorithmic}[1]
		\REQUIRE ~~\
		The channels $\{ \mathbf{G}_k(\mathbf{s})|\mathbf{s} \in {\mathcal{ S}}_{\mathrm{ T}}, k \in \mathcal{U} \}$;
		\ENSURE ~~\
		The sum rate $ R_{\mathrm{sum}} $ of the DUs, the combiners $\bm{\psi}:= \{ \bm{\psi}_k| k \in \mathcal{U}  \}$ and the H-IDET beamformer $\boldsymbol{\Theta}:= \{ \bm{\theta}_k(\mathbf{s})|\mathbf{s} \in {\mathcal{ S}}_{\mathrm{ T}}, k \in \mathcal{U} \} $;
		\STATE Initialise $ \bm{\bar{\psi}}$ and $ \boldsymbol{\bar{\Theta}}$ by (\ref{eq:35}) and (\ref{eq:36}), respectively;
		\REPEAT
		\STATE Update $ \bm{\rho}$ by (\ref{eq:16});
		\STATE Update $ \bm{\psi}$ by (\ref{eq:13}) and (\ref{eq:15});
				\STATE Initialise $ \boldsymbol{\bar{w}} $ by solving (P7);		
		\REPEAT
				\STATE Obtain $ \boldsymbol{w} $ by solving (P6);
				\STATE Caculate $ R_{\mathrm{eq}} $ by (\ref{eq:object4});
		\UNTIL $ R_{\mathrm{eq}} $ converges
		\STATE Update $ \boldsymbol{\Theta}$ by (\ref{eq:23});	
		\STATE Caculate $ R_{\mathrm{sum}} $ by (\ref{eq:object1});
		\UNTIL $ R_{\mathrm{sum}} $ converges
		\STATE Return $ R_{\mathrm{sum}} $, $ \bm{\Psi}$ and $ \boldsymbol{\Theta}$. 
	\end{algorithmic}
\end{algorithm}

\subsubsection{Convergence and Complexity Analysis}
We first briefly analyze the convergence of our proposed algorithm. According to the equivalence between $ R_{\mathrm{eq}} $ and $ R_{\mathrm{sum}} $, $ R_{\mathrm{eq}} $ is bounded since the transmit power is limited. Besides, the updates of combiner $ \bm{\Psi}^{(i+1)} $ in (\ref{eq:13}), (\ref{eq:15}) and auxiliary variable $ \bm{\rho}^{(i+1)}) $ in (\ref{eq:16}) are monotonous, which is expressed as
\begin{align}\label{eq:33}
	R_{\mathrm{eq}}(\boldsymbol{w}^{(i)},\boldsymbol{\Theta}^{(i)},\bm{\Psi}^{(i+1)},\bm{\rho}^{(i+1)})&\leq R_{\mathrm{eq}}(\boldsymbol{w}^{(i)},\boldsymbol{\Theta}^{(i)},\bm{\Psi}^{(i)},\bm{\rho}^{(i+1)}) \nonumber \\ &\leq R_{\mathrm{eq}}(\boldsymbol{w}^{(i)},\boldsymbol{\Theta}^{(i)},\bm{\Psi}^{(i)},\bm{\rho}^{(i)}).
\end{align}
Moreover, in the $ (i+1) $-th iteration of the BCD algorithm, $ \boldsymbol{w}^{(i+1)} $ is obtained based on the SCA, in which the locally tight approximations of the original problem (P5) are obtained and solved by successively imposing tight convex restrictions on the constraint sets. The converged $ \boldsymbol{w}^{(i+1)} $ is the KKT points of the originnal Problem (P5) \cite{Razaviyayn2014SuccessiveCA}, thus we have 
\begin{align}\label{eq:34}
	R_{\mathrm{eq}}(\boldsymbol{w}^{(i+1)},\boldsymbol{\Theta}^{(i)},\bm{\Psi}^{(i+1)},\bm{\rho}^{(i+1)})\leq R_{\mathrm{eq}}(\boldsymbol{w}^{(i)},\boldsymbol{\Theta}^{(i)},\bm{\Psi}^{(i+1)},\bm{\rho}^{(i+1)}).
 \end{align}
Since $ \boldsymbol{\Theta}^{(i+1)} $ is obtained by the finite-item approximation with $ \boldsymbol{w}^{(i+1)} $, the convergence of $ \boldsymbol{\Theta}^{(i+1)} $ is asymptotic. When the number of Fourier bases $ N_F $ is sufficiently large, we have 
\begin{align}\label{eq:42}
	R_{\mathrm{eq}}(\boldsymbol{w}^{(i+1)},\boldsymbol{\Theta}^{(i+1)},\bm{\Psi}^{(i+1)},\bm{\rho}^{(i+1)})\leq R_{\mathrm{eq}}(\boldsymbol{w}^{(i+1)},\boldsymbol{\Theta}^{(i)},\bm{\Psi}^{(i+1)},\bm{\rho}^{(i+1)}).
\end{align}
Combining (\ref{eq:33}), (\ref{eq:34}) and (\ref{eq:42}), $ R_{\mathrm{eq}} $ is monotonically nonincreasing. Therefore, our proposed algorithm will achieve the convergence gradually.

Then, we analyze the complexity of our proposed algorithm, which is mainly dependent on the updates of $ \boldsymbol{w} $. The sampling number of the integral $ \int _{\cal S_{\mathrm{T}}} $ is denoted by $ I_s $. Generally, $ I_s \gg K+L $ and $ N_F \gg K+L $, since the H-IDET transmitter is nearly continuous. Firstly, the complexity for initializing $ \bm{\bar{\psi}}$ and $ \boldsymbol{\bar{\Theta}}$ is $ \mathcal{O}(LI_s) $, according to (\ref{eq:35}) and (\ref{eq:36}). Then, the complexity for updating $ \bm{\rho}$ is $ \mathcal{O}(K(K+L)I_s) $, according to Eq. (\ref{eq:11}) and Eq. (\ref{eq:16}). Besides, the complexity for updating $ \bm{\psi}$ is $ \mathcal{O}((K+L)^2I_s) $ according to Eq. (\ref{eq:13}) and Eq. (\ref{eq:15}). Moreover, $ \boldsymbol{w} $ is initialized by solving (P7) with SCA, in which the convex approximation (P8) can be approximated as a second-order cone program (SOCP) problem. There are $ 6N_F $ real variables and $ L +1 $ SOC constraints in the SOCP problem. Specifically, one constraints have $ 6N_F+2 $ real dimensions while the other $ L $ constraints all have $ 2 $ real dimensions. According to \cite{LOBO1998193}, the complexity of each turn to solve the SOCP is $ \mathcal{O}((6N_F)^{2}(6N_F+2L+2)) $, while the number of required iterations is $ \mathcal{O}(\sqrt{L+1}) $. Similarly, when we update $\boldsymbol{w}$ by SCA,  there are $ 6(K+L)N_F $ real variables and $ L +2 $ SOC constraints in the approximated SOCP problem. Specifically, two constraints have $ 6(K+L)N_F+2 $ real dimensions while $ L $ constraints have $ 2 $ real dimensions. Then the complexity of each turn to solve the SOCP is $ \mathcal{O}((6(K+L)N_F)^{2}(12(K+L)N_F+2L+4)) $ and the number of required iterations is $ \mathcal{O}(\sqrt{L+2}) $. Finally, the complexity of updating $\boldsymbol{\Theta}$ is $ \mathcal{O}((K+L)I_sN_F) $ according to (\ref{eq:23}). Let $ I_1 $ denote the number of iterations required by the convergence of sum-rate $ R_{\mathrm{sum}} $. The overall complexity of \textit{Algorithm I} is approximated as $ \mathcal{O}((K+L)I_1I_sN_F+\sqrt{L+2}I_1(K+L)^3N_F^{3}) $. 

\subsubsection{Spatial orthogonality analysis}
	In this part, we will further discuss the spatial orthogonality in {\it Lemma 2}. For simplicity, we assume that the H-MIMO transmitter is linear along the $z$-direction, satisfying $  \left\lbrace \mathbf{s}=(0,0,s) \in \mathcal{C}| -c/2<s<c/2\right\rbrace  $. The user is located at $ \mathbf{s}=(d,0,r)$.  This model corresponds to single-polarized linear antennas, while the Green function of scalar radiation near-field can be  derived as 
\begin{align}
	G(\mathbf{r},\mathbf{s})= {\mathbf{G}}(\mathbf{r},\mathbf{s})(3,3)=\frac{\mathsf{i} \kappa Z_0 d^2 e^{\mathsf{i} \kappa \sqrt{(r-s)^2+d^2}}}{4\pi((r-s)^2+d^2)^{\frac{3}{2}}}.
\end{align}
Let two user are located at $ (d_1,0,r_1) $ and $ (d_2,0,r_2) $, respectively, where $ d_1\neq d_2$, $r_1 \neq r_2$. Then, the correlation function can be calculated as 
\begin{align}
	&\left|\int_{\mathcal{C}}G^H(\mathbf{r_1},\mathbf{s})G(\mathbf{r_2},\mathbf{s}) \mathrm{d}s \right| \nonumber \\=& \frac{\kappa^2 Z_0^2 d^4}{16\pi^2}\left|\int_{\mathcal{C}}\frac{e^{\mathsf{i} \kappa\left(  \sqrt{(r_1-s)^2+d_1^2}-\sqrt{(r_2-s)^2+d_2^2}\right)}  }{((r_1-s)^2+d_1^2)^{\frac{3}{2}}((r_2-s)^2+d_2^2)^{\frac{3}{2}}} \mathrm{d}s \right| .
\end{align}
\begin{figure*}[h]
	\centering
	\includegraphics [width=166mm]{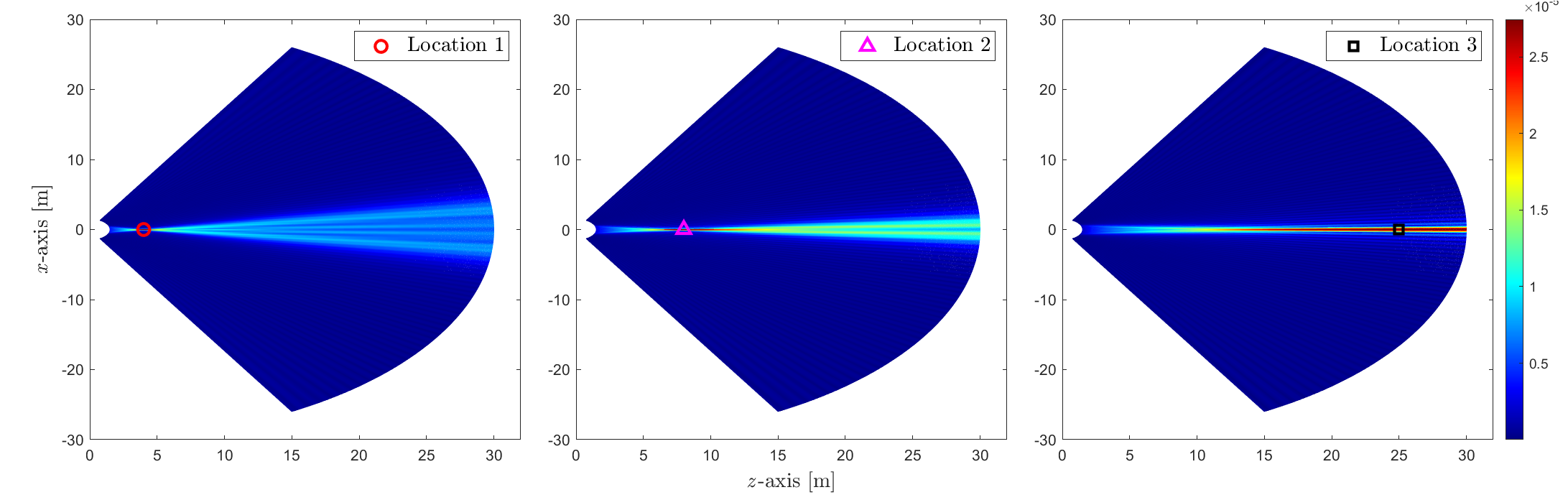}\\
	\caption{Normalized beam pattern for a single EU at different locations} 
	\label{nearfield_single_user_beam}
\end{figure*}
\begin{figure}[t]
	\centering
	\includegraphics [width=88mm]{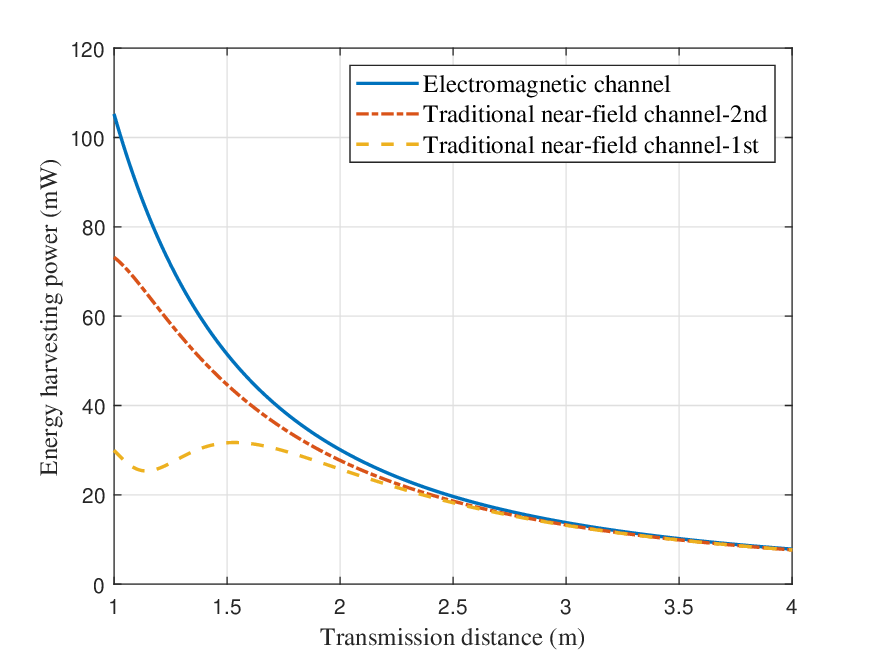}\\
	\caption{Comparison with traditonal near-field channel} 
	\label{comparison with traditonal near-field channel}
\end{figure}
We assume the users are located at the Fresnel region. The following two approximations can be adopted to phase and amplitude, respectively \cite{1137900}.
\begin{equation}
	\sqrt{(r-s)^2+d^2}\approx d+\frac{(r-s)^2}{2d},
\end{equation}
\begin{equation}
	(r-s)^2+d^2\approx r^2+d^2.
\end{equation}
Then, we have
	\begin{align}
		&\left|\int_{\mathcal{C}}G^H(\mathbf{r_1},\mathbf{s})G(\mathbf{r_2},\mathbf{s}) \mathrm{d}s \right| \nonumber \\ =& \frac{\kappa^2 Z_0^2 d^4 e^{\mathsf{i} \kappa \left( d_1-d_2+\frac{r_1^2}{2d_1}-\frac{r_2^2}{2d_2}\right)  }}{16\pi^2(r_1^2+d_1^2)^{\frac{3}{2}}(r_2^2+d_2^2)^{\frac{3}{2}}}\left|\int_{\mathcal{C}}e^{\mathsf{i} \kappa \left( \frac{s^{2}-2r_1s}{2d_1}-\frac{s^2-2r_2s}{2d_2} \right) }  \mathrm{d}s \right|
		\nonumber \\ =& \epsilon_0 \left| \int_{\mathcal{C}}e^{\mathsf{i}\left( \epsilon_1 s+ \epsilon_2 s^2\right) } \mathrm{d}s \right| 
	\end{align}
	where $ \epsilon_0=\frac{\kappa^2 Z_0^2 d^4 e^{\mathsf{i} \kappa \left( d_1-d_2+\frac{r_1^2}{2d_1}-\frac{r_2^2}{2d_2}\right)  }}{16\pi^2(r_1^2+d_1^2)^{\frac{3}{2}}(r_2^2+d_2^2)^{\frac{3}{2}}} $, $ \epsilon_1=\frac{r_2}{d_2}-\frac{r_1}{d_1} $ and $ \epsilon_2= \frac{1}{2d_1}-\frac{1}{2d_2} \neq 0 $. By substituting $ s+\frac{\epsilon_1}{2\epsilon_2} $ with $ t $, we can obtain 
	\begin{equation}
		\lim_{c \rightarrow \infty } \frac{1}{c} \left|\int_{\mathcal{C}}G^H(\mathbf{r_1},\mathbf{s})G(\mathbf{r_2},\mathbf{s}) \mathrm{d}s \right|=\lim_{c \rightarrow \infty }\frac{\epsilon_0}{c} \left| \int_{-\infty}^{\infty}e^{\mathsf{i} \epsilon_2 t^2 } \mathrm{d}t \right|.
	\end{equation}
	According to the Fresnel function \cite{1137900}, the limitation of the integral can be expressed as
	\begin{equation}
		\int_{-\infty}^{\infty}e^{\mathsf{i} \epsilon_2 t^2 } \mathrm{d}t=2\int_{0}^{\infty} \left( \cos (\epsilon_2 t^2)+\mathsf{i}\sin(\epsilon_2 t^2)  \right)  \mathrm{d}t = \sqrt{\frac{\pi}{2\epsilon_2}}(1+j).
	\end{equation}
	Therefore, we can obtain
	\begin{align}
		\lim_{c \rightarrow \infty } \frac{1}{c} \left|\int_{\mathcal{C}}G^H(\mathbf{r_1},\mathbf{s})G(\mathbf{r_2},\mathbf{s}) \mathrm{d}s \right|=	\lim_{c \rightarrow \infty } \frac{1}{c} \sqrt{\frac{\pi}{\epsilon_2}} =0, \nonumber\\ \mathrm{for}  \ d_1\neq d_2, r_1 \neq r_2,
	\end{align}
	which completes the proof of the spatial orthogonality for single-polarized linear antennas. Similar discussions can be made for other cases such as multi-polarization and planar arrays. Besides, the  speed of the convergence is affected by the locations of users. These discussions are beyond the scope of this paper and reserved for our future research. 
	
	The asymptotically EM wave received by the $ k $-th DU can be derived as
	\begin{align} 
		\lim_{| \mathcal{S}_{\mathrm{ T}} | \rightarrow \infty } \mathbf{y}_k = \underbrace{x_k \int_{\mathcal{S}_{\mathrm{ T}}} \mathbf{G}_k(\mathbf{s}) \bm{\theta}_k(\mathbf{s})\mathrm {d}\mathbf{s}}_{\varrho}+o(\varrho)+\mathbf{n}_k,
	\end{align}
	where $ f(y)\sim o(y) $ indicates that $ \lim_{y \rightarrow \infty }(f(y)/y)=0 $. Therefore, the inter-user interference asymptotically approach to zero and the sum-rate saturates to the interference-free sum-rate.

\section{Numerical Results}
In this section, we aim to evaluate the performance of our H-IDET system with the aid of simulation results, where the near-field focusing for single EU and the multi-user H-IDET are discussed, respectively.
\subsection{Near-field focusing for single EU}
Firstly, Numerical results of near-field focusing for single EU are provided in this subsection. The transmitter is deployed on the $ xy $-plane with $ {\cal S_{\mathrm{T}}}: = \left \{{ { {\left ({{s_{x},{s_{y}},{s_{z}}} }\right)} {\Big |}\left |{ {s_{x}} }\right | \le \frac {L_{x}}{2},\left |{ {s_{y}} }\right | \le \frac {L_{y}}{2},{s_{z}} = 0} }\right \} $, where $L_x=1.5\mathrm{m}, L_y=0.5 \mathrm{m} $, and the single EU is located at the z-axis. The transmit power is set to 0.003 $ \mathrm{A}^2 $ and the frequency is 10GHz. The characteristic impedance $ Z_0 $ of EM channel and that $ Z $ of energy harvester  is set to 376.73 $ \Omega $ and 25 $ \Omega $, respectively. The integral $ \int _{\cal S_{\mathrm{T}}} $'s sampling number is set to $ I_s $ = 1200.

We investigate the effect of user location on the planar beampattern in Fig. \ref{nearfield_single_user_beam}. The variables of each subfigure are different user locations, which are (0,0,4), (0,0,8) and (0,0,25), respectively. Therefore, the beam is also focused at different locations, respectively. To eliminate the effect of the distance, the harvested power at each location is normalized according to the path loss. In conventional far-field communication systems, the beam can only concentrate its energy at a certain angle, but not at different distances. Different from far-field beamsteering, the proposed scheme can achieve near-field focusing. Observe from Fig. \ref{nearfield_single_user_beam} that although the EU is located at the same angle to the antenna array, the beam can focus the power at different distance. That is because the EM channel model without far-field approximation can reveal the most essential characteristics of the channel, by means of which we can obtain resolutions of both the angular and the distance domain in the near field.

We characterize the advantages of EM channel over the traditional near-field channel in Fig. \ref{comparison with traditonal near-field channel}. Based on the Fresnel approximation \cite{1137900}, the traditional channel model is obtained under the first-order and second-order Taylor series expansion $ \sqrt{1+x}=1+x/2-x^2/8+\mathcal{O}(x^3) $, respectively. Observe from Fig. \ref{comparison with traditonal near-field channel} that the EU can harvest more energy with the aid of EM channel and the gap narrows as the distance increases. This is because the traditional channel model is inaccurate outside the lower bound of the Fresnel region, while the EM channel model is still effective. The error arisen from traditional channel model gradually shrinks as the lower bound is approached.

\begin{figure}[t]
	\centering
	\includegraphics [width=88mm]{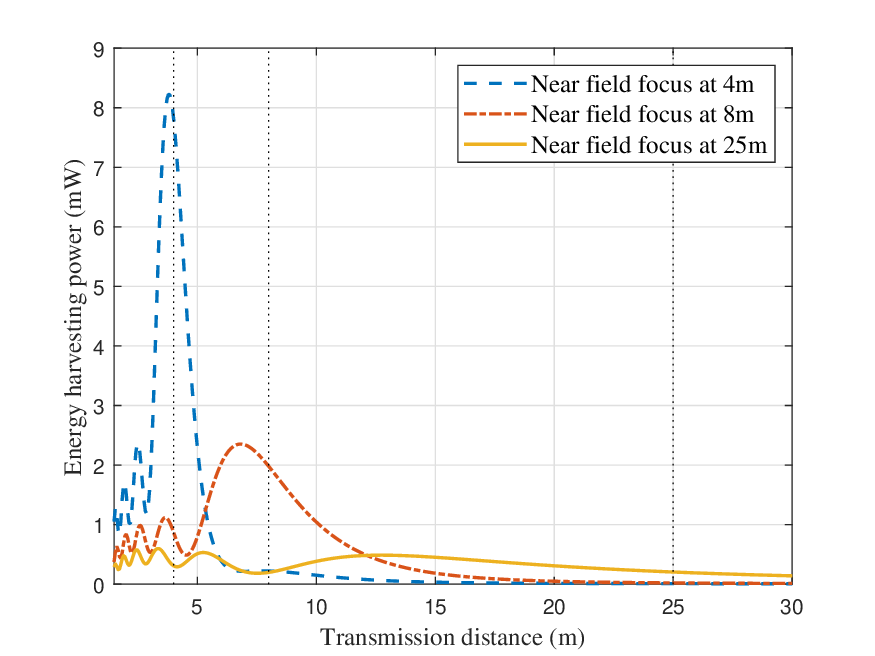}\\
	\caption{Energy harvesting power at the EU versus transmission distance.} 
	\label{Harvested energy}
\end{figure}
 We characterize the impact of the distance between the EU and transmitter on WET performance in Fig. \ref{Harvested energy}. Observe from Fig. \ref{Harvested energy} that the EU harvests the maximum energy when beams focus at its position compared to the other beams. That is because the beam is narrow not only in angular but also distance domain. The energy focusing effect is obvious among different locations, which can provide more DoF to improve the IDET performance. Besides, the closer to the transmitter, the better the focusing effect. Since the spherical wave tends to be more similar to the plane wave with the farther distance, the near-field focusing effect becomes worse, when we increase the transmission distance. On the other hand, although the beamforming gain is the highest at the focal point, the practical harvested energy is also  affected by the path-loss, thus there is a offset between the focal position and where the maximum energy harvesting power is achieved. 
 

\subsection{Multi-user H-IDET system}
Then, we evaluate the performance of the H-IDET system with the aid of holographic beamforming design in this subsection. 
The H-IDET transmitter serves $ K=4 $ DUs and $ L=2 $ EUs simultaneously. Follow the setup in \cite{10158997}, the transmitter is deployed on the $ xy $-plane where $L_x=L_y=0.3 \mathrm{m} $. The DUs are located at $ (\pm5\mathrm{m},\pm5\mathrm{m},30\mathrm{m}) $, while the EUs are located at $ (\pm1\mathrm{m},1\mathrm{m},1\mathrm{m}) $, respectively. Without specific statement, the working frequency of the H-IDET system is $ f=10 \mathrm{GHz} $ while the characteristic impedance $ Z_0 $ of channel and that $ Z $ of energy harvester is set to 376.73 $ \Omega $ and 25 $ \Omega $ respectively. The maximum transmit power and  the minimum required energy harvesting power of the EUs are set to $ P_t=0.01 \mathrm{A}^2 $ and $ \mathrm{P}_0= 1\mathrm{mW} $ respectively. The noise power is set to $ \sigma^2=5.6\times10^{-3} \mathrm{V^2/m^2}$. We assume the direction of the Poynting vector is perpendicular to the plane of the EU's receiver ,\textit{i.e.}, $ \phi=0 $. According to  \cite{ 7934322}, the circuit parameters of energy harvester are $ a=1500, b=0.0022, M=3.9\mathrm{mW} $. The integral $ \int_{\cal S_{\mathrm{T}}} $'s sampling number is set to $ I_s $ = 1024 while the Fourier base number is set to $ N_F=441 $. 

\begin{figure}[t]
	\centering
	\includegraphics [width=88mm]{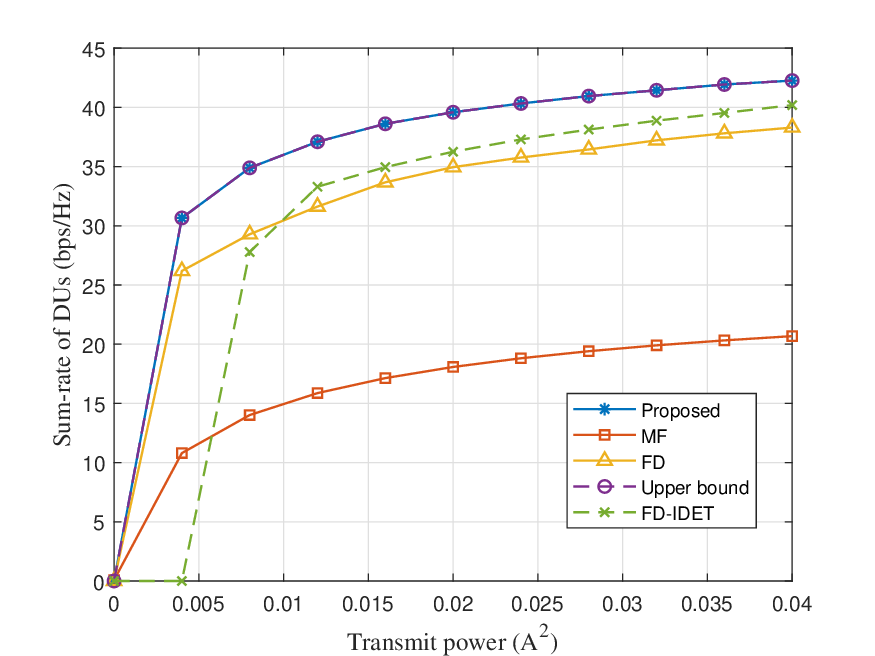}\\
	\caption{Sum-rate of DUs versus transmit power.}
	\label{fig_txpower}
\end{figure}
\begin{figure}[t]
	\centering
	\includegraphics [width=88mm]{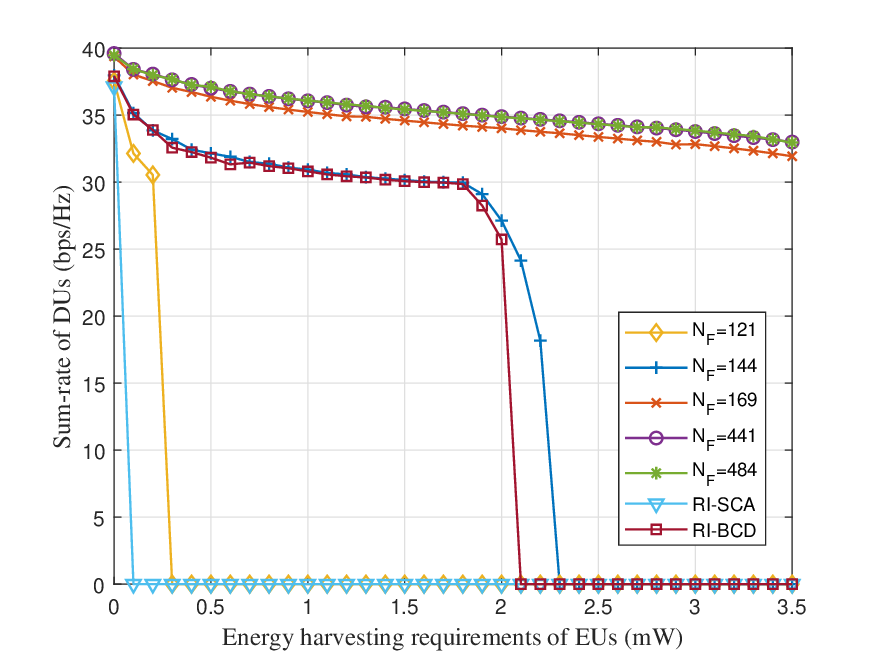}\\
	\caption{Sum-rate of DUs versus energy harvesting requirements of EUs.}
	\label{fig_energy}
\end{figure}
\begin{figure}[t]
	\centering
	\includegraphics [width=88mm]{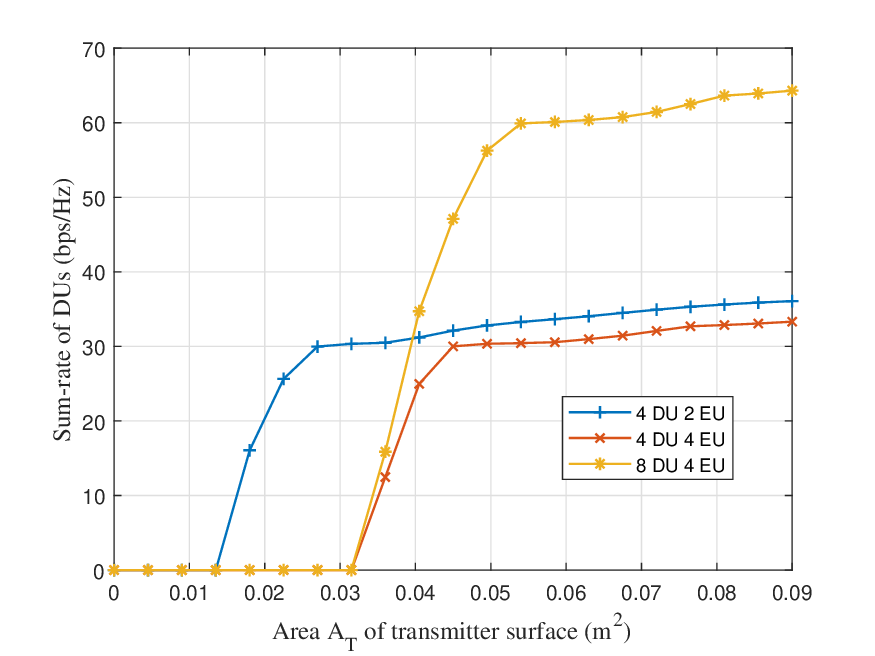}\\
	\caption{Sum-rate of DUs versus area of transmitter surface  $ \mathrm{A_T} $.} 
	\label{fig_A_T}
\end{figure}
For the performance comparison, we consider the following benchmark schemes:
\begin{figure*}
	\centering  
	\subfigbottomskip=1pt 
	\subfigcapskip=-2pt 
	\subfigure[]{
		\begin{minipage}[b]{0.48\textwidth}
			\includegraphics[width=0.495\linewidth]{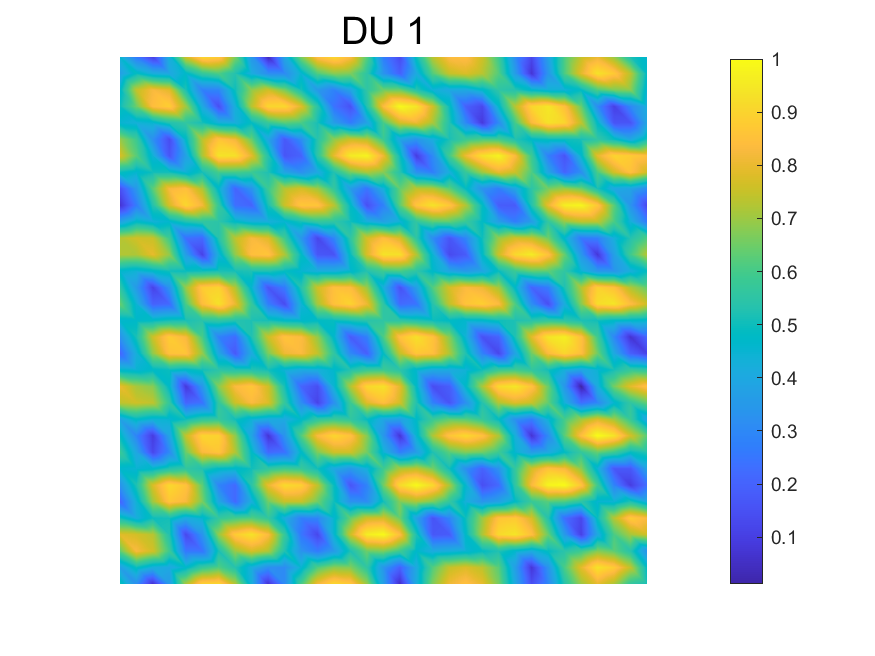} 
			\includegraphics[width=0.495\linewidth]{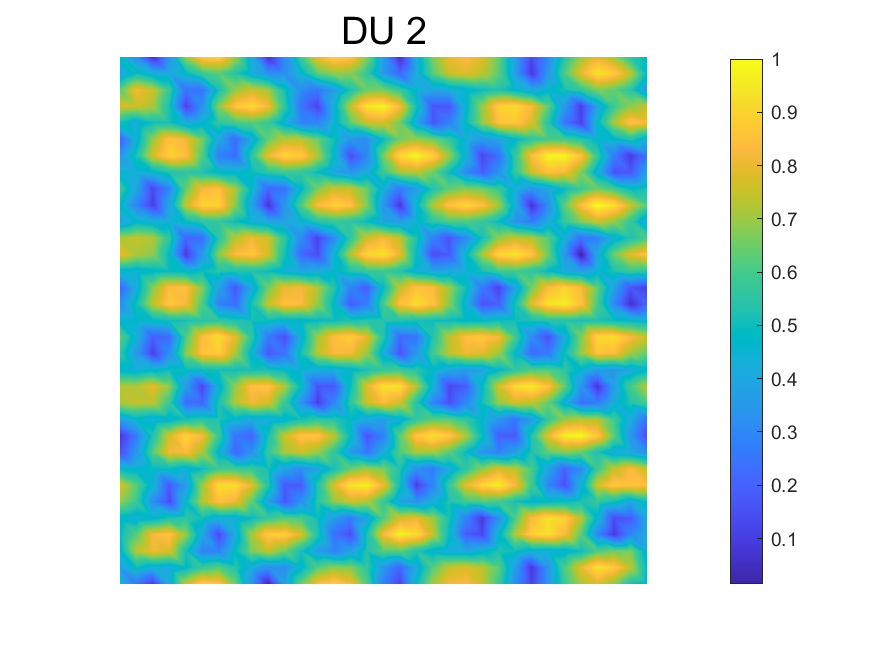}
			\\ 
			\includegraphics[width=0.495\linewidth]{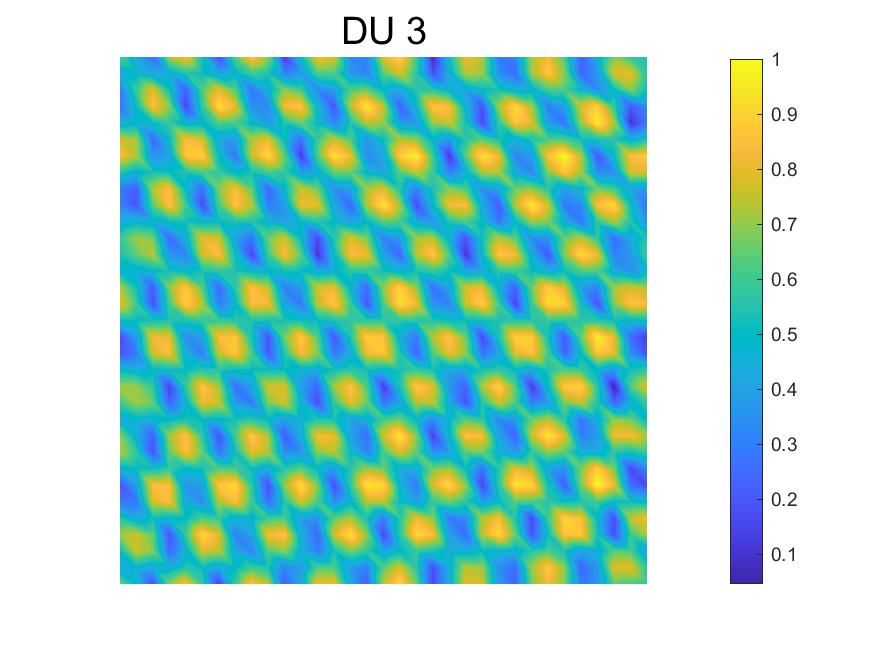}
			\includegraphics[width=0.495\linewidth]{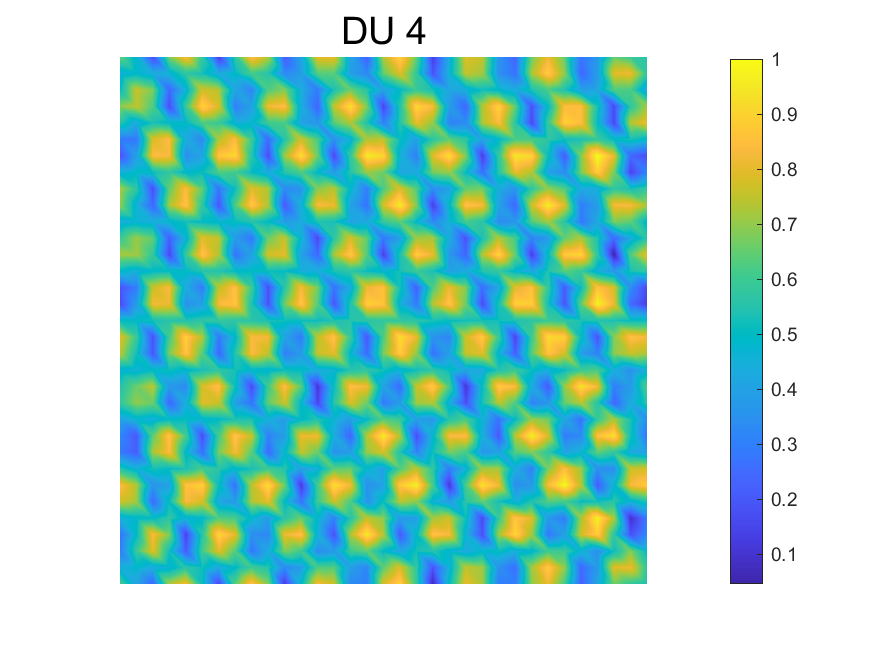}
		\end{minipage}
	}
	\subfigure[]{
		\begin{minipage}[b]{0.48\textwidth}
			\includegraphics[width=0.495\linewidth]{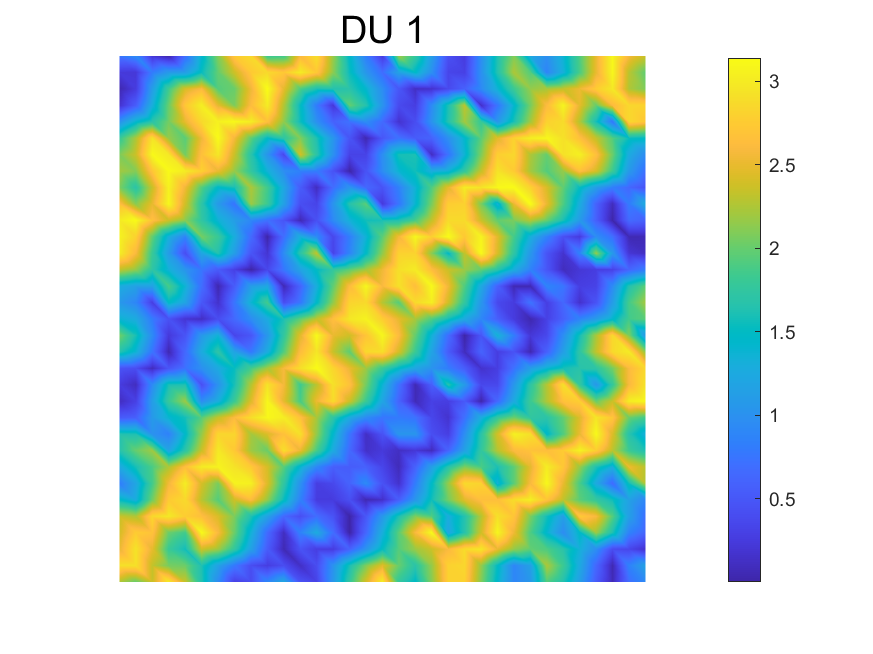}
			\includegraphics[width=0.495\linewidth]{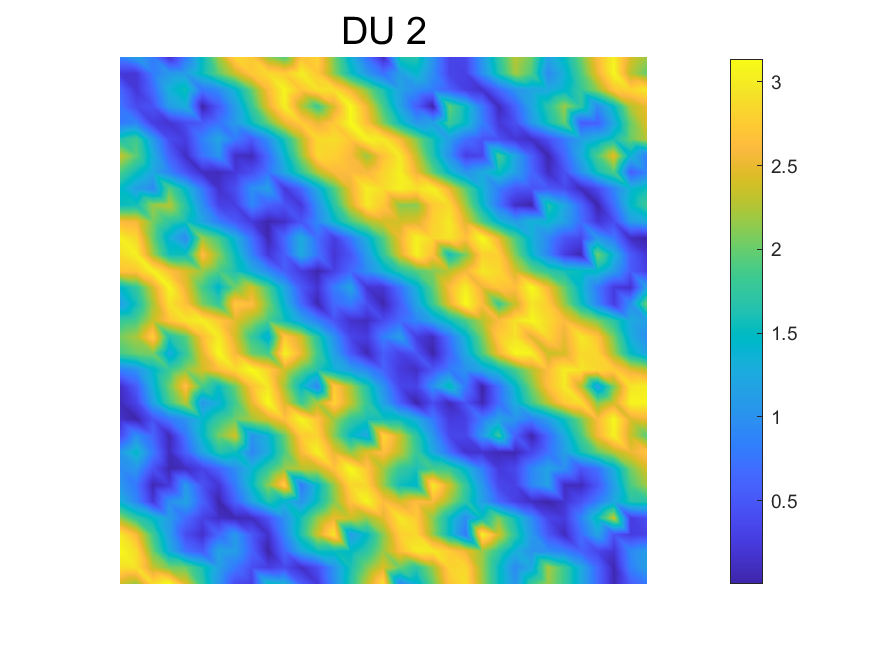}
			\\
			\includegraphics[width=0.495\linewidth]{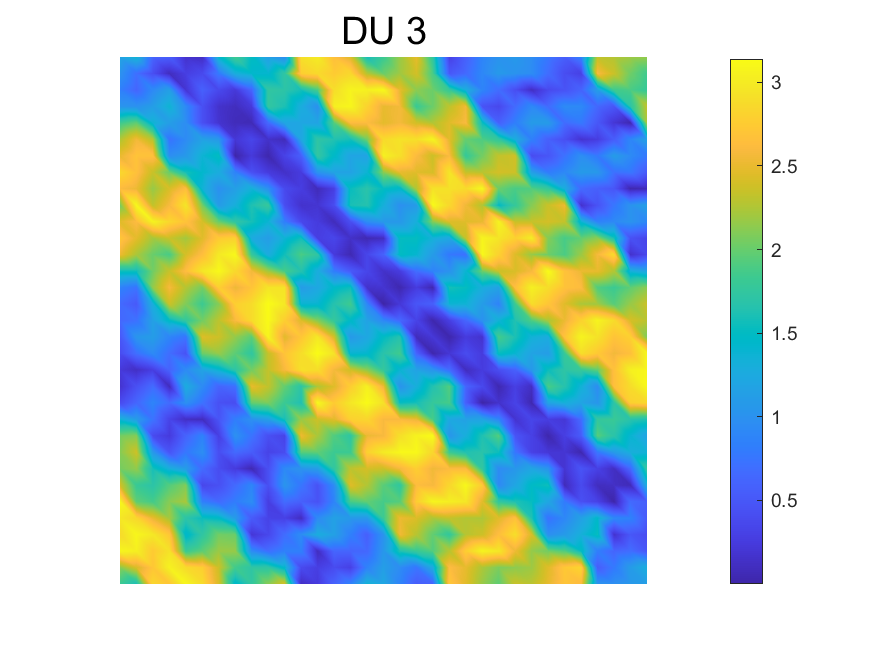}
			\includegraphics[width=0.495\linewidth]{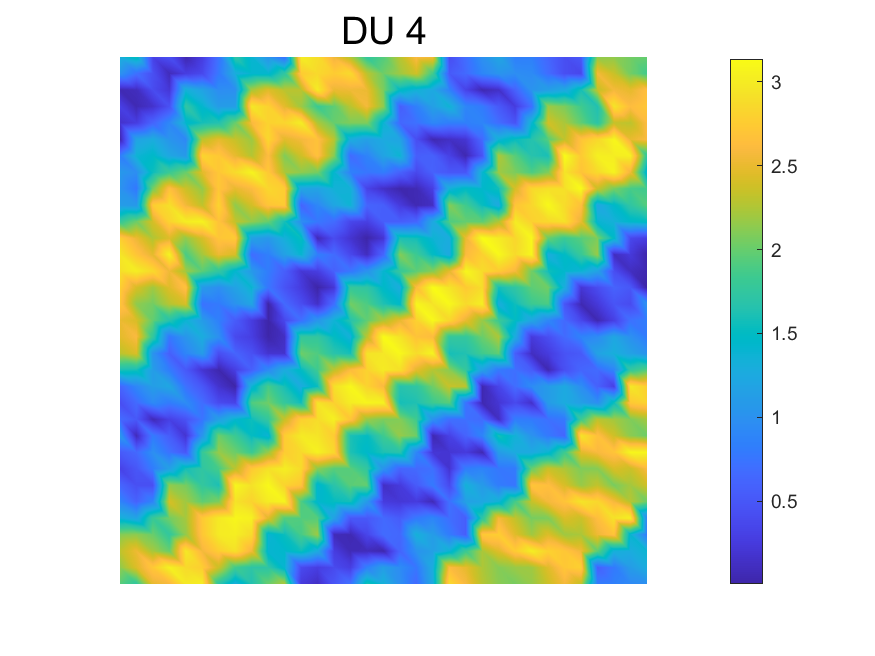}
		\end{minipage}
	}		
	\\
	\subfigure[]{
		\begin{minipage}[b]{0.48\textwidth}		
			\includegraphics[width=0.495\linewidth]{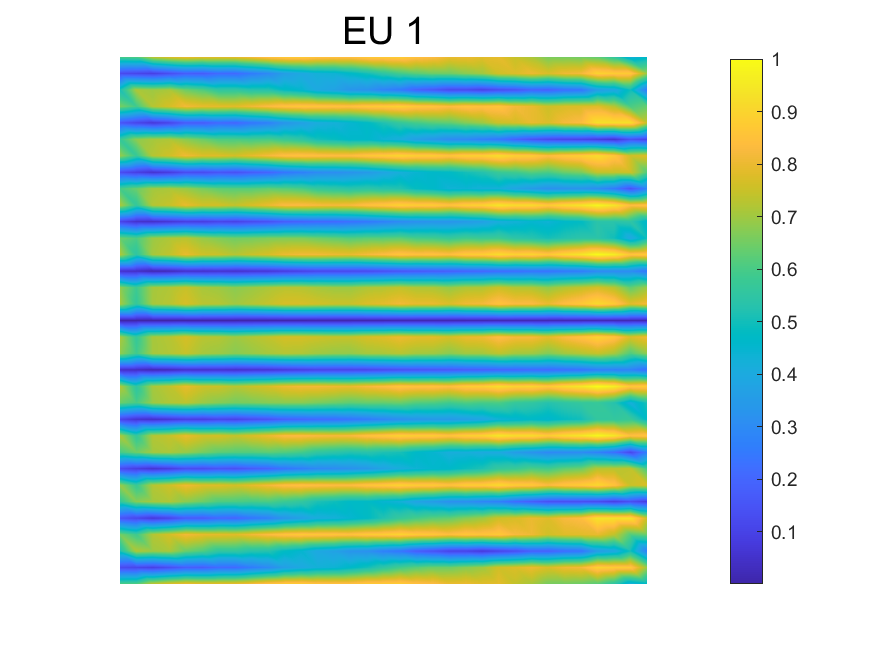}
			\includegraphics[width=0.495\linewidth]{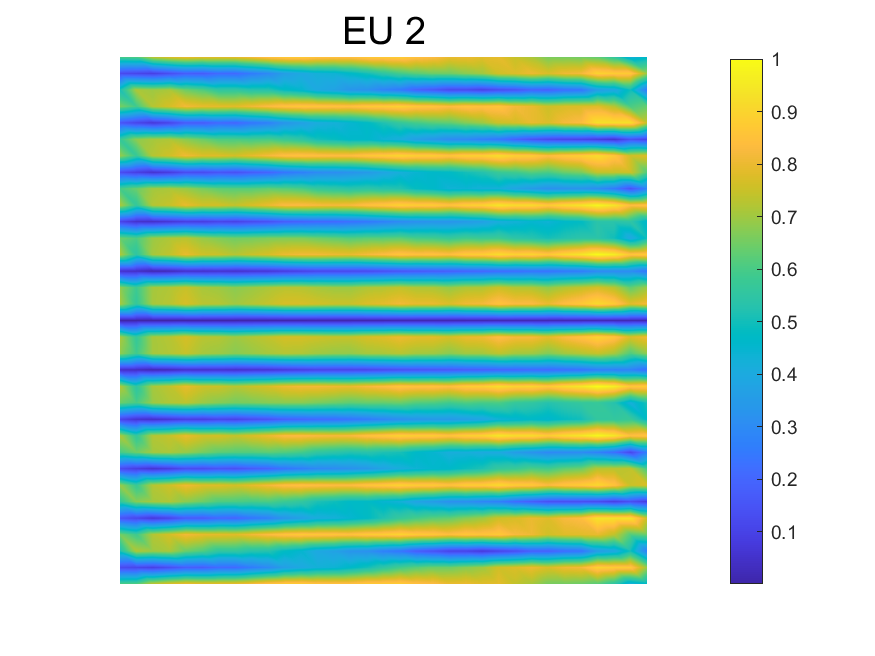}
		\end{minipage}
	}	
	\subfigure[]{
		\begin{minipage}[b]{0.48\textwidth}
			\includegraphics[width=0.495\linewidth]{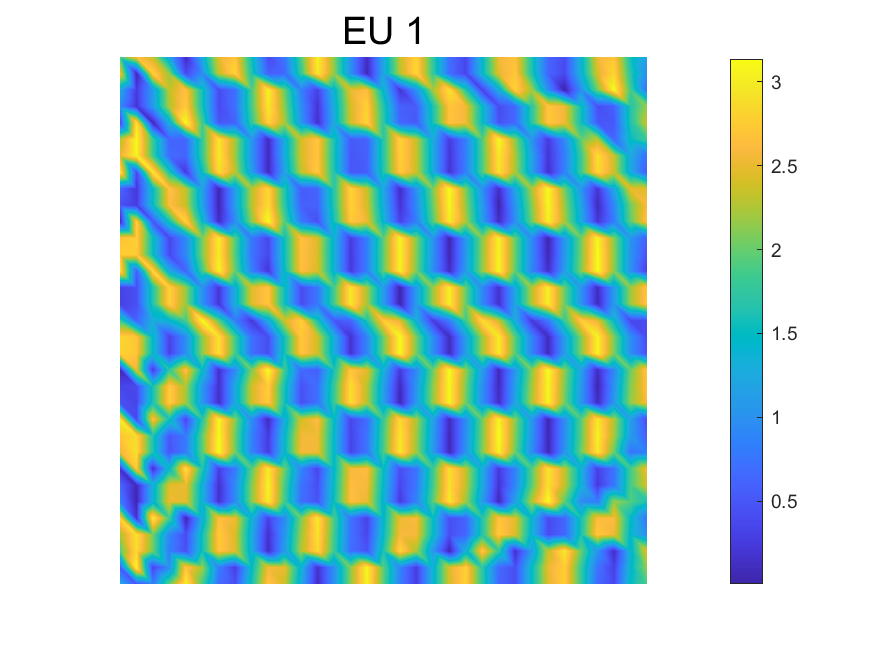}
			\includegraphics[width=0.495\linewidth]{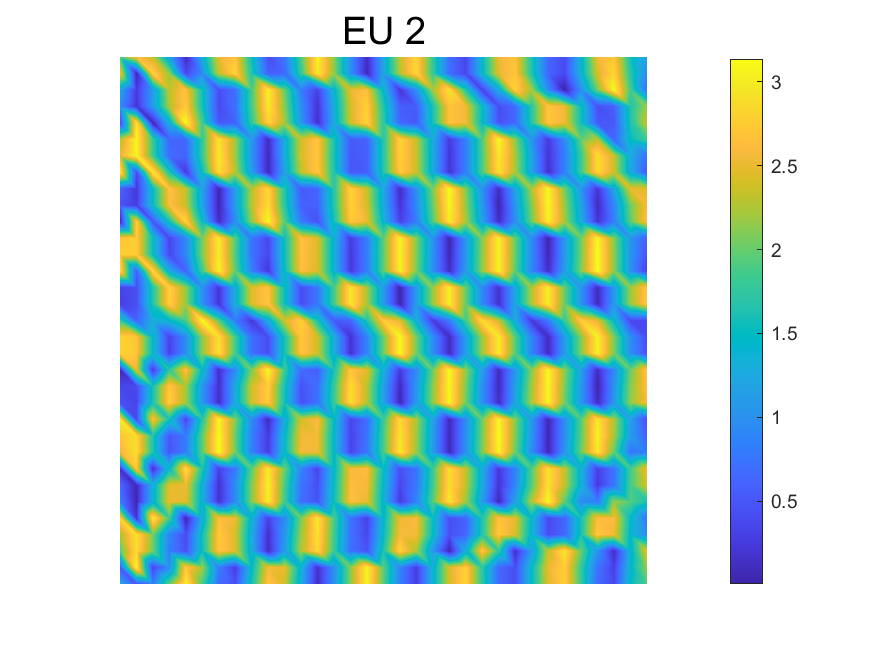}
	\end{minipage}}
	\caption{Amplitude of H-IDET for DUs (a) and EUs (c). Phase of H-IDET for DUs (b) and EUs (d).}
	\label{fig_H_MIMO}
\end{figure*}

\begin{figure*}
	\centering  
	\subfigbottomskip=1pt 
	\subfigcapskip=-2pt 
	\subfigure[]{
		\includegraphics[width=0.445\linewidth]{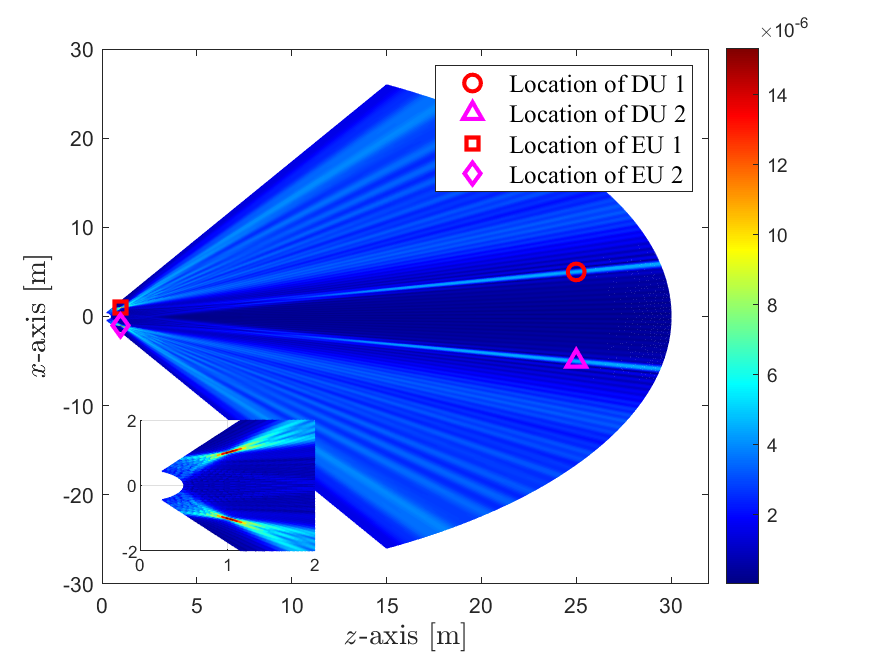} 
	}
	\subfigure[ ]{
		\includegraphics[width=0.445\linewidth]{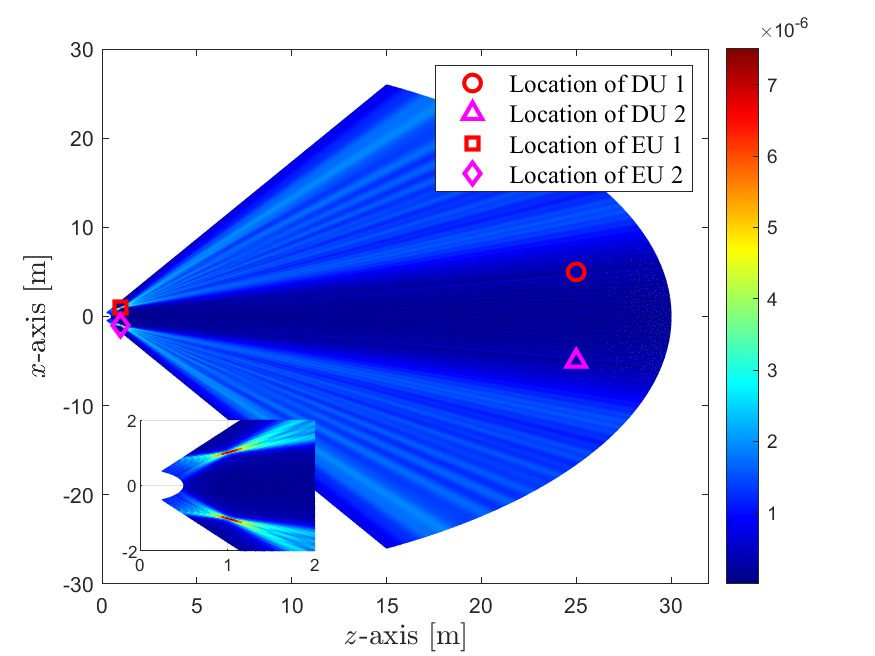} 
	}		
	\\
	\subfigure[]{
		\includegraphics[width=0.445\linewidth]{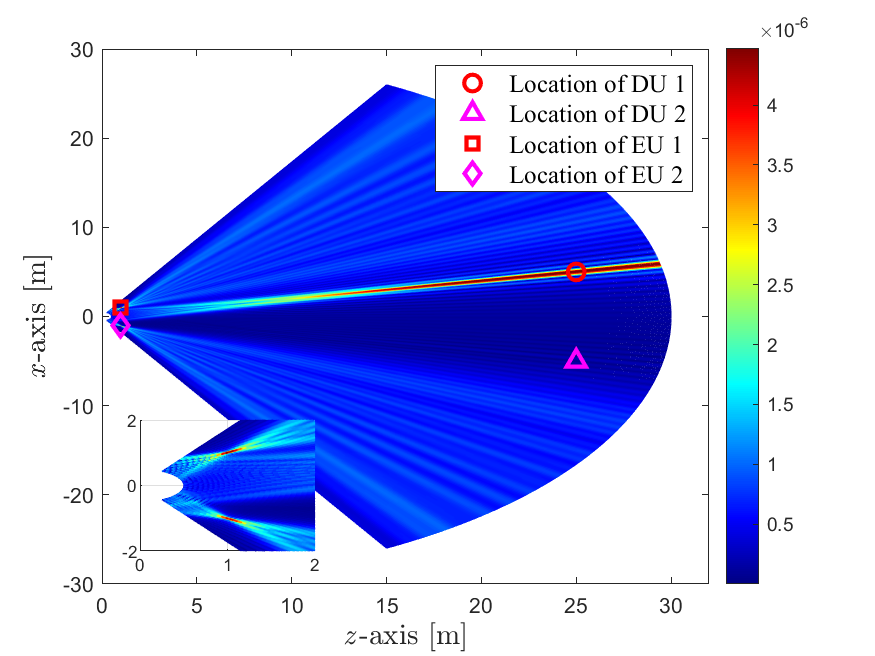} 
	}	
	\subfigure[]{
		\includegraphics[width=0.445\linewidth]{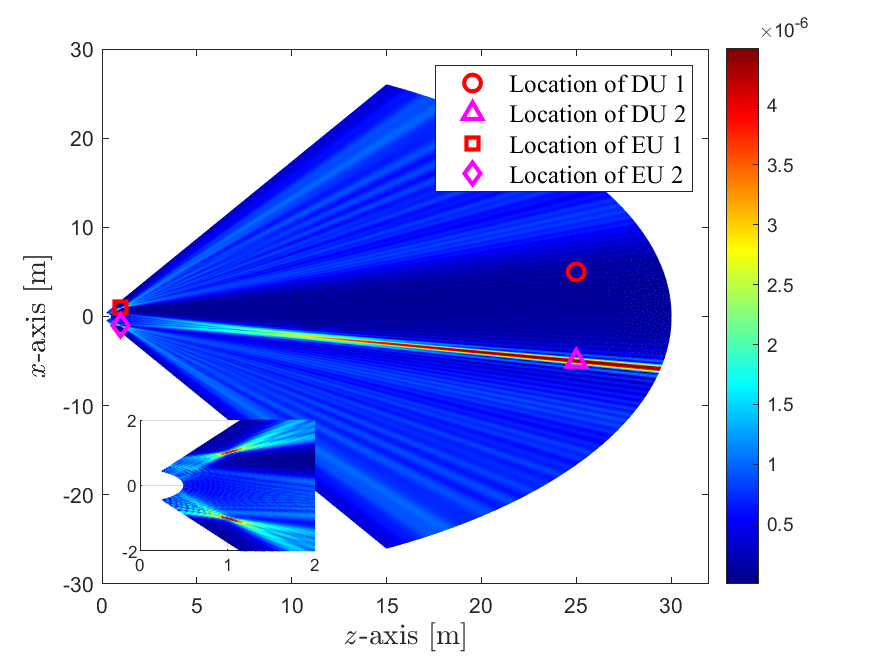} 
	}
	\caption{Near-filed focusing for the IDET system: Near-filed focusing for the IDET system: the normalized beam pattern of the whole H-MIMO (a), and the normalized beam pattern of the beamforming for EUs (b), DU 1 (c) and DU 2 (d). }
	\label{Nearfiled IDET}
\end{figure*}
\begin{itemize}
	\item {\bf Traditional fully digital (FD) MIMO:} Inspired by the comparisons between H-MIMO and traditional MIMO in \cite{9906802} and \cite{10158997}, we discretize the continuous surface of H-MIMO to represent a traditional discrete antenna array. Specifically, each discrete antenna is separated by half a wavelength, thus there are $ M = \lceil\dfrac{2L_x}{\lambda}\rceil \times\lceil\dfrac{2L_y}{\lambda}\rceil $ antennas in total. Besides, the $ m $-th discrete antenna occupies the region ${\mathcal{S}}_{m}= \left\lbrace (x,y,z)\big| |x-s_{m,x}|^2+ |y-s_{m,y}|^2 \leq \frac{A_m}{\pi},z=0\right\rbrace $, where $ (s_{m,x},s_{m,y},0) $ is the center position and the effective aperture area is assumed to be $ A_m= \lambda^2/(8\pi) $. Moreover, the beamforming for the $ k $-th user at the $ m $-th discrete tri-polarization antenna can be expressed as  $\bm{\theta}_{k,m}(\mathbf{s})=\mathrm{rect}({\mathbf{s}} \in {\mathcal{S}}_m)\mathbf{v}_{k,m} \in \mathbb{C}^{3},$
	where $ \mathrm{rect}({\mathbf{s}} \in {\mathcal{S}}_m)=\left\{
	\begin{aligned}
 1, \ & {\mathbf{s}} \in {\mathcal{S}}_m \\
 0, \ & {\mathbf{s}} \notin {\mathcal{S}}_m 
	\end{aligned}
	\right. $ is the rectangular function and $ \left\lbrace \mathbf{v}_{k,m}\right\rbrace_{m=1}^{M}  $ is the digital beamformer for user $ k $. By treating the EUs as DUs, a sum-rate maximization problem with transmit power constraints can be readily solved by the fractional programming \cite{8314727}, which is denoted by ``FD''. Besides, a sum-rate maximization problem with energy harvesting constraints of EUs can be also solved by our framework in {\textbf Algorithm I}, which is denoted by ``FD-IDET''. Both ``FD-IDET'' and ``FD'' don't use the Fourier transform in the proposed scheme.

	\item {\bf Match-filtering (MF) H-MIMO:} Xiong \emph{et al.} \cite{9139999} proposed a ``MF'' scheme to maximize the desired signal for each user in an uplink H-MIMO system. We extend this scheme to the downlink case and treat DUs and EUs as the same. The beamforming vectors for the $ k $-th user are set to $ \bm{\theta}_k(\mathbf{s})=\mu_k\mathbf{G}^H_k(\mathbf{s})\bm{\psi}_k , \mathbf{s} \in {\mathcal{ S}}_{\mathrm{ T}}$, wherein $ \bm{\psi}_k=[0,1,0]^T $ under the assumption that the antennas at users are in the y-axis polarization and $ \mu_k =\sqrt{\frac{P_t\int_{\mathcal{S}_{\mathrm{T}}} \left\| \mathbf{G}_k^H(\mathbf{s})\bm{\psi}\right\|^2\mathrm{d}\mathbf{s}}{\sum_{k=1}^{K+L}\int_{\mathcal{S}_{\mathrm{T}}} \left\| \mathbf{G}_k^H(\mathbf{s'})\bm{\psi}\right\|^2\mathrm{d}\mathbf{s'}}}$ is the normalizing factor to satisfy the power constraint. 
	
	\item {\bf Upper bound:} The sum-rate of the ``Upper bound'' is caculated by our proposed scheme with $ N_F=441 $ under ideal assumption that the interference at each user is fully canceled, i.e., the achievable rate is caculated by SNR rather than SINR.
\end{itemize}

We first characterise the impact of the transmit power versus sum-rate in Fig. \ref{fig_txpower}. Observe from Fig. \ref{fig_txpower} that the sum-rate increases, when we increase the transmit power. Besides, our H-IDET scheme outperforms traditional 'FD' and 'FD-IDET' schemes. Although a continuous aperture surface has the same DoF with discrete antenna array for the same aperture size, the performance upper bound of the system is difficult to achieved in complex scenarios with multi-user IDET and polarized electromagnetic waves. Continuous aperture surface can manipulate electromagnetic waves more flexibly over the entire plane and approximate the performance upper bound better than discrete arrays. Moreover, with the aid of H-IDET, the amplitude and the phase of the current over the entire plane are optimized, while traditional discrete MIMO can only manipulate discrete areas. Therefore, the discrete beamforming remains unchanged in the region of each discrete antenna, which causes mismatches with the channel that varies with position. Furthermore, our IDET scheme outperforms the schemes of 'MF' and 'FD'. This is because in 'MF' and 'FD' schemes, there is no efficient power allocation among EUs and DUs, which further degrades the IDET performance. Moreover, our scheme achieves the similar performance to the interference-free ``Upper Bound''. By carefully designing the beamforming of H-IDET, our scheme is able to achieve efficient interference cancellation in the simulated scenario.

We investigate the trade-off between the sum-rate of DUs and the energy harvesting power by EUs in Fig. \ref{fig_energy}. Observe from Fig. \ref{fig_energy} that the sum-rate declines when EUs has a stringent energy harvesting requirement. That's because more transmit power is allocated to EUs, which results in a reduction of sum-rate of DUs. Besides, a larger $N_F$ results in a higher IDET performance, since the Fourier base number $ N_F $ represents the approximation accuracy in Eq. (\ref{eq:18}). When $ N_F $ is not less than $ (2\lceil\frac{L_x}{\lambda}\rceil+1)(2\lceil\frac{L_y}{\lambda}\rceil+1)(2\lceil\frac{L_z}{\lambda}\rceil+1)=441 $, there are sufficient functional degrees of freedom for approximation \cite{9906802}. Moreover, ``RI-SCA'' and ``RI-BCD'' denote the random initialization instead of our proposed scheme in the step (1) and step (5) of Algorithm I, respectively. $ N_F $ is set to 144 in ``RI-SCA'' and ``RI-BCD'', respectively. It is observed that the sum-rate of random initialization suddenly drops to 0 as the energy harvesting requirement of EUs increases. This is because the random initialization results in no feasible solution to the approximated convex problem. Since the initialization of $ \boldsymbol{\bar{w}} $ has the most influence to iterations, ``RI-SCA'' drops much early than ``RI-BCD'', although ``RI-BCD'' has a lower approximation accuracy. The robustness of our algorithm is significantly improved by our proposed initialization scheme. 

We characterise the impact of transmitter's surface area $ \mathrm{A_T} $ versus the sum-rate in Fig. \ref{fig_A_T}. Observe from Fig. \ref{fig_A_T} that the sum-rate increases with a larger antenna aperture, since more spatial gains can be achieved. Besides, more DUs or less EUs  both result in a higher sum-rate, since more EUs will compete for more transmit power while more DUs will provide a higher diversity gain for the system. Moreover, the x-intercept of the curve increases with a lager number of EUs, Since a  larger antenna aperture is required to satisfy higher energy harvesting constraints.


Finally, we investigate the optimized current distributions of our beamforming design. The normalized amplitude and the phase of the x-component of the optimized beamformers $ \{\bm{\theta}_k(\mathbf{s}) \}_{k=1}^{K+L} $ are presented in Fig. \ref{fig_H_MIMO}, respectively. Observe from Fig. \ref{fig_H_MIMO} (a) that each DU carefully utilizes different regions of the H-IDET, which induces that the electric field at the position of DUs are orthogonal to each other and achieves a higher sum-rate. Then, observe from Fig. \ref{fig_H_MIMO} (b) that the phase of each DU's beamformer are nearly symmetrically distributed, which steers the electromagnetic waves toward both the DUs and EUs. Moreover, observe from Fig. \ref{fig_H_MIMO} (c) and (d) that the EUs have almost the same beamformers, since different WET beamforming have the same impacts for different EUs and can readily improve their WET performance, which yields  non-orthogonal electric fields at the EUs' position.

We then analyse the near-field focus effect for the H-IDET system in Fig. \ref{Nearfiled IDET}. Generally, the near-field range is related to the equivalent array width. Without loss of generality, we set $L_x=1.5\mathrm{m}, L_y=0.03 \mathrm{m} $ to reduce the computational complexity of simulation. Two DUs are located at $ (\pm5,0,25) $ while two EUs are located at $ (\pm1,0,1) $. The received power at each location is normalized by the path loss. Observe from Fig. \ref{Nearfiled IDET}(a) that the integrated H-IDET is able to focus energy at the position of EUs and DUs respectively. Specifically, the beam pattern of EUs is illustrated in Fig. \ref{Nearfiled IDET}(b), where the beamforming of EUs only focuses beams at the EUs and achieves the spatial filtering effect for the DUs. Moreover, the beam pattern of DUs are illustrated in Fig. \ref{Nearfiled IDET}(c) and Fig. \ref{Nearfiled IDET}(d), respectively, where the beams are focused at their target DU. At the same time, the beams also point to the EUs for better WET performance and achieve the spatial filtering effect for other DUs.

\section{Conclusions}
In this paper, we investigated a novel H-IDET system, where the H-MIMO technique is utilized to properly achieve energy focusing and eliminate inter-user interference with the aid of its strong EM manipulation capability. A BCD-based algorithm has been proposed to optimize the beamforming for maximizing the DUs’ sum-rate by guaranteeing the EUs' energy harvesting requirements. Numerical results illustrated  that both near-field focusing and spatial filtering can be implemented by our proposed scheme. Besides, the sum-rate of our scheme is able to achieve the upper bound, which is higher than other benchmark schemes. Furthermore, thank to its strong EM control capability and continuous aperture, our proposed H-IDET system achieves a better IDET performance than traditional MIMO assisted IDET system.

 \bibliography{IEEEtran}
\bibliographystyle{ieeetr}	
	
\begin{IEEEbiography}[{\includegraphics[width=1in,height=1.25in,clip,keepaspectratio]{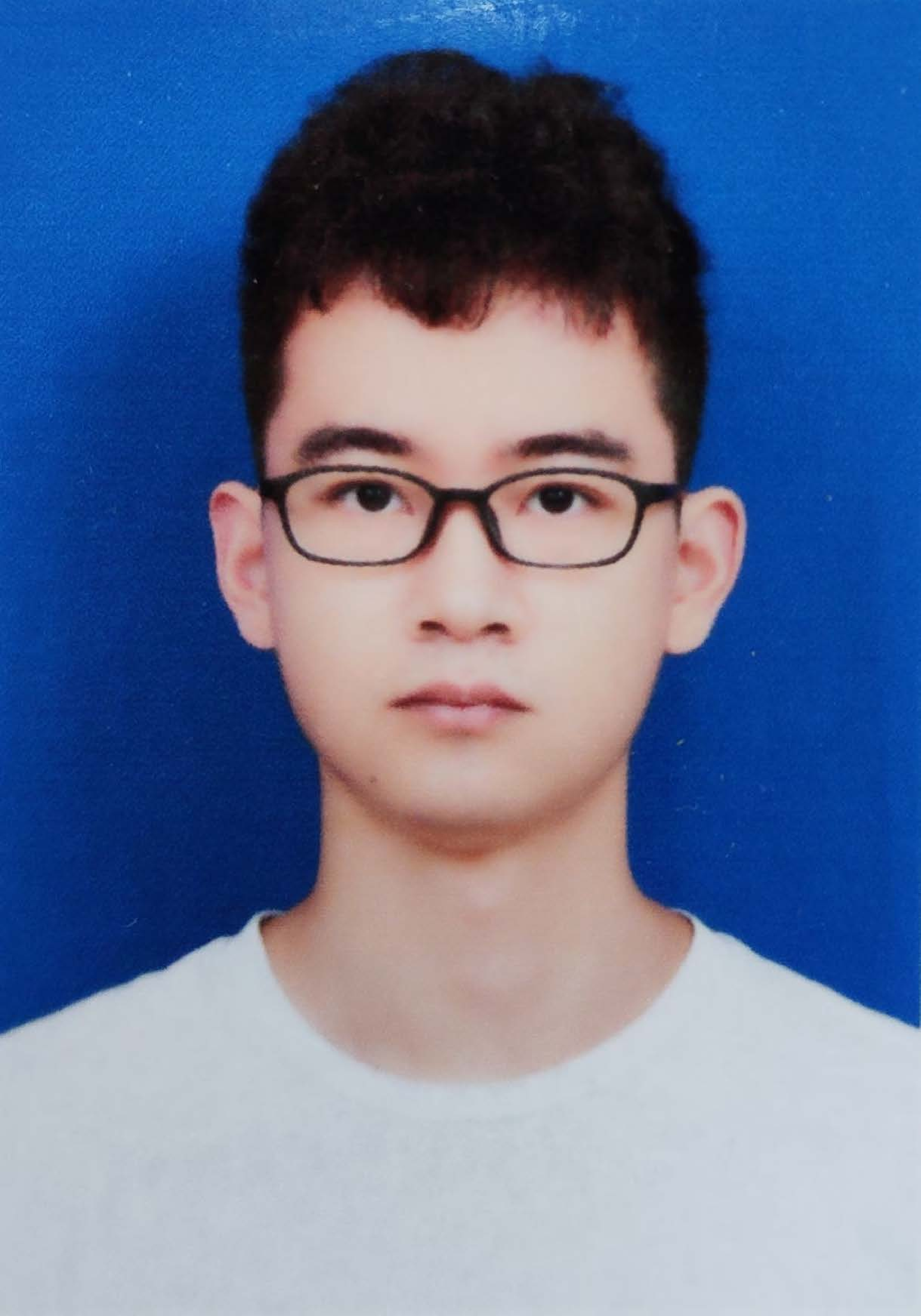}}]{Qingxiao Huang}[S'24]
	received his B.Eng. degree from Xidian Univesity, China, in 2022. He is currently working toward the M.Sc. degree at University of Electronic Science and Technology of China (UESTC). His research interests include holographic MIMO, near-field communication, intelligent reflect surface and simultaneous wireless information and power transfer.
\end{IEEEbiography}
\begin{IEEEbiography}[{\includegraphics[width=1in,height=1.25in,clip,keepaspectratio]{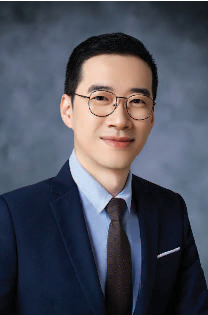}}]{Jie Hu}
	[S'11, M'16, SM'21] (hujie@uestc.edu.cn) received his B.Eng. and M.Sc. degrees from Beijing University of Posts and Telecommunications, China, in 2008 and 2011, respectively, and received the Ph.D. degree from the School of Electronics and Computer Science, University of Southampton, U.K., in 2015. Since March 2016, he has been working with the School of Information and Communication Engineering, University of Electronic Science and Technology of China (UESTC). He is now a Research Professor and PhD supervisor. He won UESTC's Academic Young Talent Award in 2019. Now he is supported by the ``100 Talents'' program of UESTC. He is an editor for \textit{IEEE Wireless Communications Letters}, \textit{IEEE/CIC China Communications} and \textit{IET Smart Cities}. He serves for \textit{IEEE Communications Magazine}, \textit{Frontiers in Communications and Networks} as well as \textit{ZTE communications} as a guest editor. He is a technical committee member of ZTE Technology. He is a program vice-chair for IEEE TrustCom 2020, a technical program committee (TPC) chair for IEEE UCET 2021 and a program vice-chair for UbiSec 2022. He also serves as a TPC member for several prestigious IEEE conferences, such as IEEE Globecom/ICC/WCSP and etc. He has won the best paper award of IEEE SustainCom 2020 and the best paper award of IEEE MMTC 2021. His current research focuses on wireless communications and resource management for B5G/6G, wireless information and power transfer as well as integrated communication, computing and sensing.
\end{IEEEbiography}
\begin{IEEEbiography}[{\includegraphics[width=1in,height=1.25in,clip,keepaspectratio]{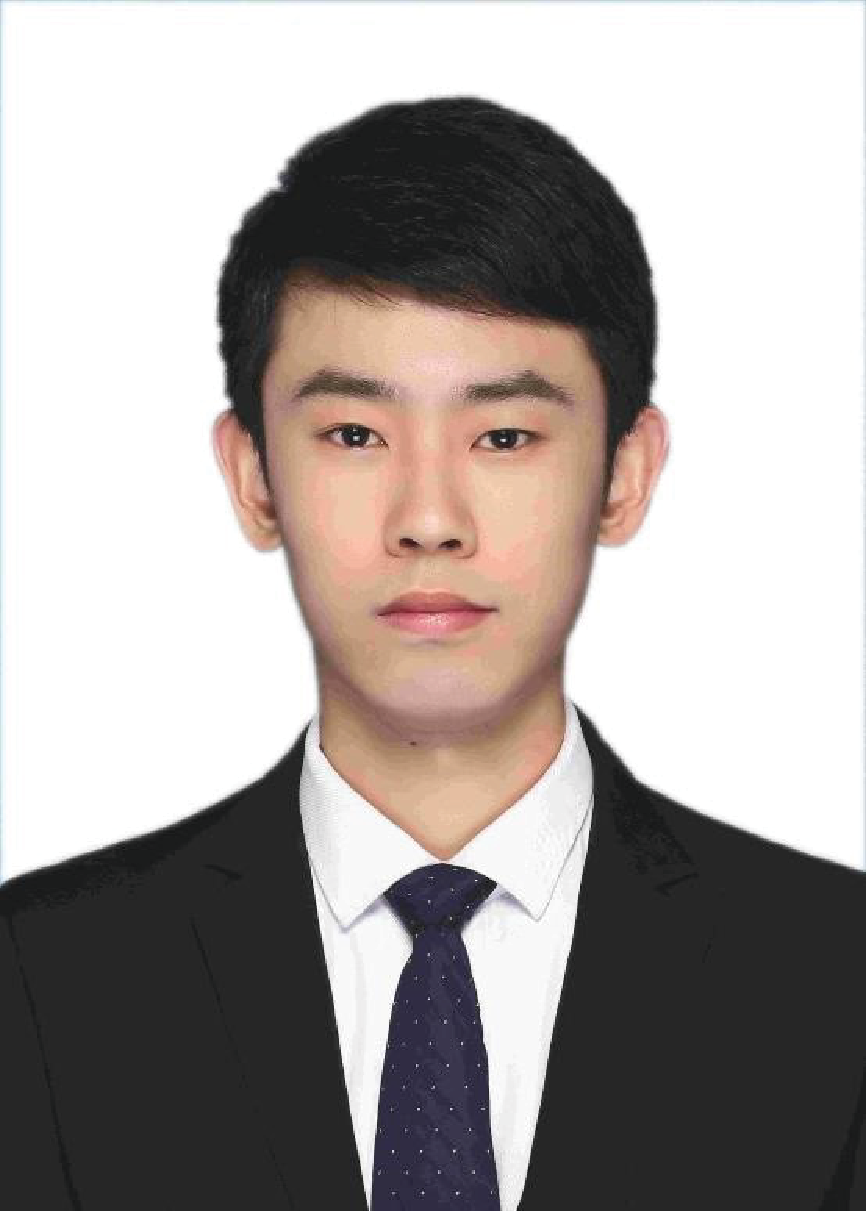}}]{Yizhe Zhao}[S'16, M'21] received the PhD in 2021 in School of Information and Communication Engineering from University of Electronic Science and Technology of China (UESTC), where he is currently an associate professor. He has been a visiting researcher with the Department of Electrical and Computer Engineering, University of California, Davis, USA. He is a member of IEEE and a senior member of China Institute of Communications. He is selected in Young Elite Scientists Sponsorship Program by China Association for Science and Technology (CAST). He serves for China Communications and Journal of Communications and Information Networks (JCIN) as the Guest Editor, and is also a TPC member of several prestigious IEEE conferences, such as IEEE ICC, Globecom. He was the recipient of IEEE CSE Best Paper Award in 2023. His research interests include modulation and coding design, integrated data and energy transfer, fluid antenna systems.
\end{IEEEbiography}
\begin{IEEEbiography}[{\includegraphics[width=1in,height=1.25in,clip,keepaspectratio]{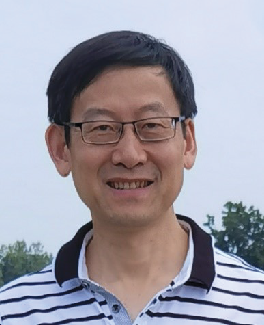}}]{Kun Yang}
	[F’22] received his PhD from the Department of Electronic \& Electrical Engineering of University College London (UCL), UK. He is currently a Chair Professor in the School of Intelligent Software and Engineering, Nanjing University, China. He is also an affiliated professor of University of Essex, UK. His main research interests include wireless networks and communications, communication-computing cooperation, and new AI (artificial intelligence) for wireless. He has published 500+ papers and filed 50 patents. He serves on the editorial boards of a number of IEEE journals (e.g., IEEE WCM, TVT, TNB). He is a Deputy Editor-in-Chief of IET Smart Cities Journal. He has been a Judge of GSMA GLOMO Award at World Mobile Congress – Barcelona since 2019. He was a Distinguished Lecturer of IEEE ComSoc (2020-2021). He is a Member of Academia Europaea (MAE), a Fellow of IEEE, a Fellow of IET and a Distinguished Member of ACM.
\end{IEEEbiography}

\end{document}